\documentclass[twocolumn]{aastex62}
\pdfoutput=1 
\usepackage{amsmath,amstext}
\usepackage[T1]{fontenc}
\usepackage{apjfonts} 
\usepackage[figure,figure*]{hypcap}
\usepackage{gensymb}
\usepackage{array}
\usepackage{graphicx}

\newcolumntype{H}{>{\setbox0=\hbox\bgroup}c<{\egroup}@{}}

\newcommand{\lv}{$l-v$}

\shorttitle{The BAaDE Survey at 86 GHz using ALMA}
\shortauthors{M.C. Stroh et al.}

\begin{document}

\accepted{August 14th, 2019}
\title{The Bulge Asymmetries and Dynamical Evolution (BAaDE) SiO Maser Survey at 86 GHz with ALMA}

\author{Michael C. Stroh}
\affiliation{Department of Physics \& Astronomy, \\The University of New Mexico, Albuquerque, NM 87131}
\author{Ylva M. Pihlstr{\"o}m}
\altaffiliation{Y.\ M.\ Pihlstr{\"o}m is also an Adjunct Astronomer at the National Radio\\ Astronomy Observatory.}
\affiliation{Department of Physics \& Astronomy, \\The University of New Mexico, Albuquerque, NM 87131}
\author{Lor\'ant O. Sjouwerman}
\affiliation{National Radio Astronomy Observatory, \\Array Operations Center, Socorro, NM 87801}
\author{Megan O. Lewis}
\affiliation{Department of Physics \& Astronomy, \\The University of New Mexico, Albuquerque, NM 87131}
\author{Mark J Claussen}
\affiliation{National Radio Astronomy Observatory, \\Array Operations Center, Socorro, NM 87801}
\author{Mark R. Morris}
\affiliation{Department of Physics \& Astronomy, \\University of California, Los Angeles, CA 90095}
\author{R. Michael Rich}
\affiliation{Department of Physics \& Astronomy, \\University of California, Los Angeles, CA 90095}
\begin{abstract}
We report on the first 1,432 sources observed using the Atacama Large Millimeter/submillimeter Array (ALMA), from the Bulge Asymmetries and Dynamical Evolution (BAaDE) survey, which aims to obtain tens of thousands of line-of-sight velocities from SiO masers in Asymptotic Giant Branch (AGB) stars in the Milky Way.
A 71\% detection rate of 86 GHz SiO masers is obtained from the infrared color-selected sample, and increases to 80\% when considering the likely oxygen-rich stars using \textit{Midcourse Space Experiment} (MSX) colors isolated in a region where $[\textnormal{D}]-[\textnormal{E}] \le 1.38$.
Based on Galactic distributions, the presence of extended CS emission, and likely kinematic associations, the population of sources with $[\textnormal{D}]-[\textnormal{E}] > 1.38$ probably consists of young stellar objects, or alternatively, planetary nebulae.
For the SiO detections we examined whether individual SiO transitions provide comparable stellar line-of-sight velocities, and found that any SiO transition is suitable for determining a stellar AGB line-of-sight velocity.
Finally, we discuss the relative SiO detection rates and line strengths in the context of current pumping models.
\end{abstract}


\section{Introduction}
The Bulge Asymmetries and Dynamical Evolution (BAaDE) survey aims to improve our understanding of the structure of the inner Galaxy and Galactic Bulge (L.\ O.\ Sjouwerman et al.\ 2019, in preparation).
By using line-of-sight velocities of SiO maser emission from red giant stars, these stars act as point-like probes of the Galactic gravitational potential.

Interstellar extinction greatly reduces the extent of the Milky Way that surveys utilizing stellar or compact probes can reach, in turn necessitating piecemeal approaches to model the spiral arms and Galactic bulge.
Since the spiral arms span large ranges of Galactic longitude, to constrain the Galactic structure models, multiple surveys must be used, often with their own limitations.
Gas emission has been much more successful in uniformly mapping larger regions of the Milky Way\citep[e.g.\ the CO($1-0$) survey by][]{2001ApJ...547..792D}; however, gas and stellar probes may lead to differences in the measured gravitational potential.
Thus a large and uniform stellar survey is required to complement the large scale gas surveys.

\textit{Infrared Astronomical Satellite} (IRAS) two-color diagrams have successfully differentiated between oxygen- and carbon-rich, evolved Asymptotic Giant Branch (AGB) populations, while also distinguishing between less evolved objects with thinner envelopes (Miras and semi-regular variables) and the thicker envelopes associated with OH/IR stars \citep{1988A&A...194..125V}. 
However, the $1'$ angular resolution of IRAS leads to confusion close to the Galactic plane, limiting its usefulness for source selection in regions closest to the Milky Way's disk.
\textit{Midcourse Space Experiment} (MSX) has an $18''$ point spread function, thus it does not suffer from the same level of source confusion throughout the Galactic plane as IRAS.
Differentiating between these numerous populations is more difficult with MSX mid-infrared colors, as the longest wavelength band is only 21.3$\mu$m, compared to the longer 60$\mu$m band of IRAS.
However, substantial progress has been made supplementing MSX colors with data in the near-infrared (see Section \ref{sec:cs_colors}).

The BAaDE survey assumes sampling of an oxygen-rich, Mira-dominated population since it is associated with region \textit{iii}a of the MSX two-color diagram, thus consisting of stars with thin circumstellar shells providing conditions favorable for SiO maser emission \citep{10.1088/0004-637X/705/2/1554}.
Additionally, both radiative and collisional excitation models suggest that SiO maser emission forms in the shells around thin-shelled AGB stars, with the maser emission radiating tangential to the shell surface.
The expected ring-like structure of the maser emission regions has been confirmed by Very Long Baseline Interferometry (VLBI) observations \citep[e.g.][]{10.1038/371395a0,2000A&A...565A.127D,10.1086/379347}, and demonstrates that SiO maser emission is usually an accurate tracer of the stellar line-of-sight velocity.
\citet{1991A&A...242..211J} found no systematic red- or blue-shifting of the SiO $v$=1 emission velocity compared to the velocities established by thermal CO and OH emission, further validating the strength of SiO maser emission as a stellar line-of-sight velocity tracer.

The full BAaDE sample consists of 28,062 sources selected by MSX colors.
All sources were selected from version 2.3 of the MSX point source catalog \citep{2003yCat.5114....0E} that coincide with MSX color region \textit{iii}a \citep{10.1088/0004-637X/705/2/1554}. 
The MSX A, C, D and E bands are centered on wavelengths of 8.28, 12.13, 14.65 and 21.34$\mu$m, respectively\footnote{Henceforth, A, C, D \& E will refer to the MSX bands, while [A], [C], [D] \& [E] will refer to the zero-magnitude corrected magnitudes for the respective band. For example $[\textnormal{D}]-[\textnormal{E}]$ refers to the difference between the zero-magnitude corrected magnitudes of the D \& E MSX bands. }. 
MSX \textit{iii}a color boundaries are given by $-0.6 \le [\textnormal{A}]-[\textnormal{D}] \le +0.4$, or $-0.7 \le [\textnormal{A}]-[\textnormal{E}] \le +0.4$ or $-0.75\le [\textnormal{C}]-[\textnormal{E}] \le +0.1$.
The MSX [A]--[D], [A]--[E], and [C]--[E] color selections correspond to 27,130 candidates, 563 candidates, and 369 candidates, respectively.
The majority (18,988) of the BAaDE sources have been surveyed with the Karl G.\ Jansky Very Large Array (VLA) using the 43 GHz ($J$=1--0) rotational transitions of SiO. 
The VLA cannot observe 9,074 BAaDE sources (32\% of the full sample) since they are located at declinations below $-35\degree$. 
This low-declination sample at $-110\degree < l < -5\degree$ contains regions of the Galactic bar furthest from us.
Some symmetry along the Galactic bar and the spiral structure may be assumed when modeling the structure of the Galactic plane, however, filling in this region with data from stellar objects using ALMA is essential for meticulous testing of the structure of the Milky Way when combined with the larger BAaDE sample observed with the VLA.
Since ALMA cannot observe the 43 GHz frequencies associated with $J$=1--0 SiO maser transitions, the ALMA sample is instead surveyed in the $J$=2--1 rotational transitions at 86 GHz.

Section \ref{sec:observations_and_data_analysis} describes the observations and spectral line detection algorithm.
Section \ref{sec:results} summarizes the line detection results.
Section \ref{sec:discussion} discusses the CS emission population, how the majority of CS emitters are likely not associated with late-type AGB stars, and how they can be filtered out of the survey.
Finally, Section \ref{sec:sio_maser_characteristics} discusses general trends of the 86 GHz SiO masers and how they relate to pumping models.

\begin{deluxetable}{lcrrl}[t]
\tablecaption{ALMA spectral window configuration\label{tab_alma_spw_config}}
\tabletypesize{\scriptsize}
\tablehead{
\colhead{} & \colhead{Central} & \colhead{Channel} & \colhead{Number} & \colhead{Potential}\\
\colhead{Spectral} & \colhead{Frequency} & \colhead{Width} & \colhead{of} & \colhead{Lines}\\
\colhead{Window} & \colhead{(MHz)} & \colhead{(km/s)} & \colhead{Channels} & \colhead{Covered}
}
\startdata
2A & 85~646.6055 & 0.855 & 1920 & SiO $v$=2, $^{29}$SiO $v$=0 \\
1   & 86~249.6630 & 0.848 & 3840  & SiO $v$=1, H$^{13}$CN \\
2B & 86~853.2095 & 0.843 & 1920 & SiO $v$=0, H$^{13}$CO+\\
3   & 97~988.0684 & 0.747 & 3840 & CS \\
\enddata
\tablecomments{The channel width is calculated at the central frequency in each spectral window.}
\end{deluxetable}

\begin{figure}[t]
\includegraphics[scale=0.425]{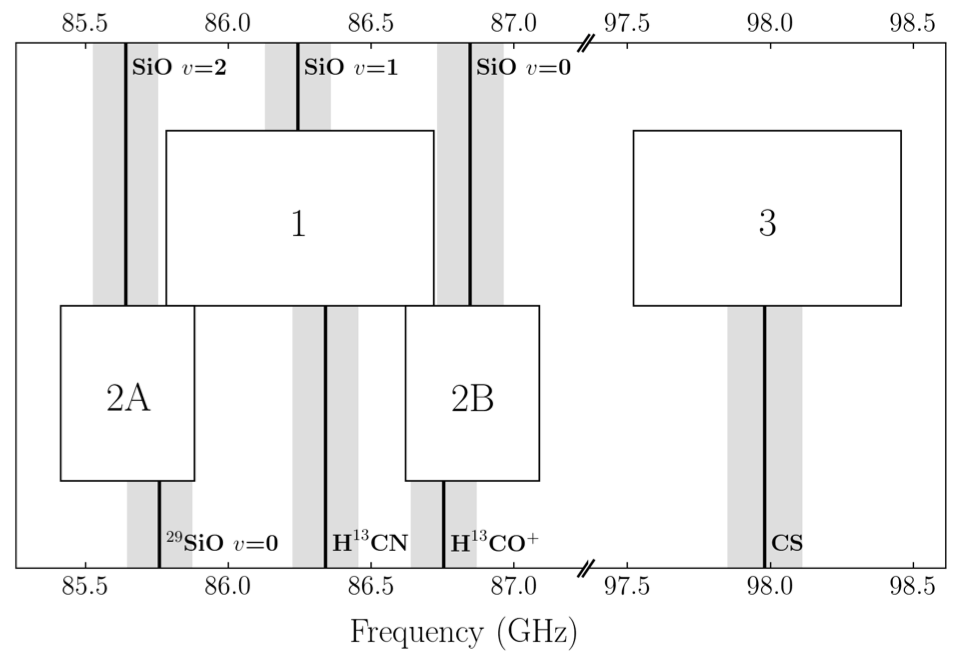}
\caption{Spectral windows for BAaDE ALMA observations. The black lines represent the rest frequency of each transition, with the gray regions representing a $\pm$400~km~s$^{-1}$ range surrounding each rest frequency.}
\label{fig_baade_alma_spws}
\end{figure}

\section{Observations and Data Analysis} \label{sec:observations_and_data_analysis}

This paper reports on the 1432 sources observed during ALMA Cycles 2, 3 and 5 (June 2014 through September 2018).
To minimize slewing overhead and enable efficient phase calibration, the sources were grouped based on their angular separation.

\subsection{Observations} \label{sec:observations}
The ALMA spectral setup in Band 3 is described in Table \ref{tab_alma_spw_config} and in Figure \ref{fig_baade_alma_spws}.
The main objective was to cover the SiO $v$=0, 1 \& 2 transitions, which also allowed coverage of the $^{29}$SiO $v$=0 isotopologue transition.
Unless otherwise noted, the SiO transitions refer to the 86 GHz $J$=2--1 transitions. 
Additionally, SiO refers to the $^{28}$SiO transitions and we will specifically reference $^{29}$SiO when discussing isotopologue transitions.
With the aim of deriving velocities from sources that instead may belong to a carbon-rich population, an additional baseband was placed on the CS ($J$=2--1) transition at 98 GHz, which has been observed, for example, in the carbon-rich archetype IRC+10216 \citep{1985A&A...147..143H}.
Several other transitions from carbon-bearing molecules can be found within our setup, most notably H$^{13}$CN and H$^{13}$CO$^{+}$.
The width of the primary beam was $\approx$1$\arcmin$.
Each channel has a width of $244.141$~kHz.

In total, 184, 1155 and 93 sources were observed during ALMA Cycles 2, 3 and 5, respectively. 
Cycle 2 observations occurred between January 17th and January 19th, 2015, with a final observing run on April 13th, 2015. 
Cycle 3 observations occurred between December 27th, 2015 and January 21st, 2016, and Cycle 5 observations were performed on January 28th, 2018.
A few runs were interrupted and/or re-observed in order to meet ALMA quality standards. 
In the case of multiple observations, results are reported for the last time a source was observed, with the exception of 16 sources where an earlier observing run resulted in additional line detections. 

As the ALMA Cycles had differing numbers of available antennas, for each cycle, the time on target per source was planned to produce an equivalent rms sensitivity of 15 mJy/beam/channel.
For the three cycles, an average rms sensitivity of 14 mJy/beam/channel was obtained.
Phase calibrators were observed with a cadence of approximately 10 minutes.

\subsection{Calibration} \label{sec:calibration}
The BAaDE ALMA calibration pipeline is based on the VLA calibration pipeline developed for the BAaDE survey, see L.\ O.\ Sjouwerman et al.\ (2019, in preparation).
However, instead of using the \textit{Astronomical Image Processing System} \citep[AIPS,][]{10.1007/0-306-48080-8_7}, we use the \textit{Common Astronomy Software Applications} \citep[CASA,][]{2007ASPC..376..127M} package.

The pipeline produces spectra for sources assumed to be at the phase center of each pointing, and the resulting signal-to-noise ratio (SNR) of a detection is therefore dependent on the accuracy of the source positions.
The BAaDE sample was selected from the MSX catalog with 1-2$\arcsec$ positional accuracy within an $18\arcsec$ beam, hence given the ALMA synthesized beam ($\approx 3\arcsec$) more precise source positions could increase the detection rate.
Especially for weaker sources, a 2$\arcsec$ positional error could result in a non-detection in the pipeline.
In the future, full imaging of the fields will be performed to find weak emission offset from the phase center, but that will require significantly more computer and time resources than currently available. 
\citet{10.3847/1538-4357/aae77d} found that for known 43 GHz SiO masers, only a $0\farcs12$ mean offset exists between the derived VLA and 2MASS positions.
Since 2MASS associations exist for 96\% of the BAaDE survey, in order to increase our detection rate, if a 2MASS association existed for a BAaDE source, the phase center was shifted to the 2MASS position using the task \textit{fixvis}.
The phase center was adjusted before any standard calibration was performed.

Because the ALMA calibrator catalog lists 91.46 and 103.49 GHz flux densities of QSO B1424--41, QSO B1921--293, and PKS 1613--586 at regular intervals, these quasars were used for flux and bandpass calibration. 
For flux calibration in our two main observing bands at 86 GHz and 98 GHz, a first order spectral slope was calculated using calibrator measurements taken on the date nearest to our observations on which both 91.46 and 103.49 GHz were available in the ALMA calibrator database.
Most observing runs had such calibrator information available within 7 days of the observations, with a maximum offset of 16 days.
91.46 GHz was used as the reference frequency for the extrapolation.
Our flux calibration, which relied upon the ALMA calibrator catalog, differed by less than 5\% from flux calibration using resolved solar system objects and the 'Butler-JPL-Horizons 2012' standard \citep{butler_memo}.
Phase calibration was performed using PKS 1613--586, PKS 1714--336, or PMN J1650--5044.

\subsection{Line Detections} \label{sec:line_detections}
In the pipeline, a line is considered to be a detection depending on the SNR.
First, an initial rms noise was calculated individually for each spectral window.
In order to remove the influence of spectral lines on the noise calculations, spectral regions exceeding 8 times the initial rms noise were removed first.
A robust rms noise was calculated from the remaining regions of the spectrum.

Single line detections were required to have an SNR above 5$\sigma$.
Multiple line detections could be made down to a 4.5$\sigma$ level, if the velocities from each line agreed within 30 km~s$^{-1}$.
This large velocity range was adopted to address the large 86 GHz $v$=1 emission extents of supergiants by \citet{10.1088/0004-6256/149/3/100}.
In the discussion on line velocity offsets in Section \ref{sec:velocities}, the SiO line-peak velocities are in much closer agreement than this 30 km~s$^{-1}$ allowance.
However, the sources with both SiO and carbon-bearing molecular transitions can have line-peak velocities in disagreement by $\approx 20$~km~s$^{-1}$ (see Section \ref{sec:compact_cs_population}).
The line finding algorithm selects only the highest SNR peak for a given transition by filtering out secondary emission peaks and/or broad line structure within 30 km~s$^{-1}$ of a brighter detected line.

For each source with one or more detections, the brightest channel with a line detection was subsequently used for phase-only self-calibration. 
For all successful self-calibrations, the phase corrections were applied to all four spectral windows. 
The self-calibration worked well for the SiO detections, but always failed when a CS line was used. 
Possibly this is due to the extended nature of most CS emission for non-stellar sources (see Section \ref{sec:two_carbon_populations}). 

Ideally, the self-calibration solutions should be transferred to neighboring source scans, thereby improving the phase calibration for weak sources and hence also the detection rate.
Indeed, this is an important part of the calibration strategy for the VLA observations  where phase calibrators typically are not readily available (L.\ O.\ Sjouwerman et al.\ 2019, in preparation).
This type of phase-transfer is not possible in CASA at this time, but as ALMA automatically includes frequent calibrator scans, and 2MASS positions are used, a high detection rate was still achieved.
After applying the self-calibration corrections, the final set of spectra was produced, along with a list of detections based on the SNR requirements outlined above.

\newpage

\begin{deluxetable*}{lrrrrrrrrrrr}[h]
\tablecaption{86 GHz BAaDE source detections\label{tab:alma_baade_detections_cycle_2_3}}
\tabletypesize{\scriptsize}
\tablehead{
\colhead{BAaDE} & \colhead{Observing Date} & \colhead{R.A.} & \colhead{Declination} & \colhead{Position} &
\colhead{S(SiO $v$=0)} & 
\colhead{S(SiO $v$=1)} & 
\colhead{S(SiO $v$=2)} & 
\colhead{S($^{29}$SiO $v$=0)} & 
\colhead{S(H$^{13}$CN)} & 
\colhead{S(H$^{13}$CO$^{+}$)} & 
\colhead{S(CS)} \\
\colhead{Source} & \colhead{(YYYY-MM-DD)} & \colhead{(J2000)} & \colhead{(J2000)} & \colhead{Catalog} &
\colhead{(Jy/channel)} & 
\colhead{(Jy/channel)} & 
\colhead{(Jy/channel)} & 
\colhead{(Jy/channel)} & 
\colhead{(Jy/channel)} & 
\colhead{(Jy/channel)} & 
\colhead{(Jy/channel)}
}
\startdata
ad3a-22252 & 2015-01-17 & $16:00:11.163$ & $-53:47:11.96$ & 2MASS & $<$   0.062  &   0.123 $\pm$   0.013 & $<$   0.066  & $<$   0.066  & $<$   0.066  & $<$   0.062  & $<$   0.077 \\
ad3a-22234 & 2015-01-17 & $16:00:13.619$ & $-53:50:41.36$ & 2MASS & $<$   0.066  &   0.649 $\pm$   0.013 & $<$   0.071  & $<$   0.071  & $<$   0.066  & $<$   0.066  & $<$   0.081 \\
ad3a-22435 & 2015-01-19 & $16:00:17.943$ & $-53:04:28.71$ & 2MASS & $<$   0.060  &   0.148 $\pm$   0.013 & $<$   0.064  & $<$   0.064  & $<$   0.063  & $<$   0.060  & $<$   0.077 \\
ad3a-22575 & 2015-01-18 & $16:00:19.705$ & $-52:31:51.83$ & 2MASS & $<$   0.058  & $<$   0.060  & $<$   0.064  & $<$   0.064  & $<$   0.060  & $<$   0.058  & $<$   0.075 \\
ad3a-22906 & 2015-01-19 & $16:00:21.013$ & $-51:14:53.85$ & 2MASS & $<$   0.063  &   0.354 $\pm$   0.013 & $<$   0.068  & $<$   0.068  & $<$   0.066  & $<$   0.063  & $<$   0.079 \\
ad3a-22600 & 2015-01-18 & $16:00:22.488$ & $-52:24:37.22$ & 2MASS & $<$   0.059  & $<$   0.061  & $<$   0.062  & $<$   0.062  &   0.062 $\pm$   0.012 &   0.129 $\pm$   0.012 &   0.570 $\pm$   0.015\\
ad3a-22398 & 2015-01-19 & $16:00:29.088$ & $-53:11:41.35$ & 2MASS & $<$   0.062  &   2.283 $\pm$   0.013 & $<$   0.066  & $<$   0.066  & $<$   0.064  & $<$   0.062  & $<$   0.077 \\
ad3a-22823 & 2015-01-19 & $16:00:30.879$ & $-51:34:33.93$ & 2MASS & $<$   0.065  &   0.360 $\pm$   0.014 & $<$   0.071  & $<$   0.071  & $<$   0.068  & $<$   0.065  & $<$   0.078 \\
ad3a-22468 & 2015-01-19 & $16:00:35.453$ & $-52:57:49.69$ & 2MASS & $<$   0.061  &   1.341 $\pm$   0.012 & $<$   0.065  & $<$   0.065  & $<$   0.062  & $<$   0.061  & $<$   0.076 \\
ad3a-22215 & 2015-01-17 & $16:00:38.881$ & $-53:55:21.66$ & 2MASS & $<$   0.064  &   0.168 $\pm$   0.013 & $<$   0.065  & $<$   0.065  & $<$   0.065  & $<$   0.064  & $<$   0.078 \\
\nodata & \nodata & \nodata & \nodata & \nodata  &   \nodata &  \nodata & \nodata  & \nodata  & \nodata  & \nodata & \nodata \\
\enddata
\tablecomments{
    Upper limits are reported at the 5$\sigma$ level across a single $244.141$~kHz channel; however, in the line detection algorithm, a 4.5$\sigma$ threshold was used if at least one other line was detected.
    The fifth column states whether the position in columns 3 and 4 is the 2MASS- or MSX-associated position.
    Table \ref{tab:alma_baade_detections_cycle_2_3} can be found in its entirety in a machine-readable format.
}
\end{deluxetable*}

\begin{deluxetable*}{lrrrrrrrrrrr}[h]
\tablecaption{86 GHz BAaDE line-peak line-of-sight velocities\label{tab:alma_baade_velocities_cycle_2_3}}
\tabletypesize{\scriptsize}
\tablehead{
\colhead{BAaDE} & \colhead{$l$} & \colhead{$b$} & 
\colhead{v(SiO $v$=0)} & 
\colhead{v(SiO $v$=1)} & 
\colhead{v(SiO $v$=2)} & 
\colhead{v($^{29}$SiO $v$=0)} & 
\colhead{v(H$^{13}$CN)} & 
\colhead{v(H$^{13}$CO$^{+}$)} & 
\colhead{v(CS)} & 
\colhead{$\langle \textnormal{v}_{\textnormal{SiO}}\rangle$} &
\colhead{$\langle \textnormal{v}_{\textnormal{C}}\rangle$} \\
\colhead{Source} & \colhead{($^{\circ}$)} & \colhead{($^{\circ}$)} & 
\colhead{(km~s$^{-1}$)} & 
\colhead{(km~s$^{-1}$)} & 
\colhead{(km~s$^{-1}$)} & 
\colhead{(km~s$^{-1}$)} & 
\colhead{(km~s$^{-1}$)} & 
\colhead{(km~s$^{-1}$)} & 
\colhead{(km~s$^{-1}$)} & 
\colhead{(km~s$^{-1}$)} & 
\colhead{(km~s$^{-1}$)}
}
\startdata
ad3a-22252 & $-31.383~523$ & $ -0.605~114$ & \nodata  & $    -81.9$ & \nodata  & \nodata  & \nodata  & \nodata  & \nodata  & $  -81.9$ & \nodata \\
ad3a-22234 & $-31.416~932$ & $ -0.653~113$ & \nodata  & $    -87.0$ & \nodata  & \nodata  & \nodata  & \nodata  & \nodata  & $  -87.0$ & \nodata \\
ad3a-22435 & $-30.905~795$ & $ -0.076~880$ & \nodata  & $    -86.1$ & \nodata  & \nodata  & \nodata  & \nodata  & \nodata  & $  -86.1$ & \nodata \\
ad3a-22906 & $-29.706~616$ & $  1.300~734$ & \nodata  & $    -99.6$ & \nodata  & \nodata  & \nodata  & \nodata  & \nodata  & $  -99.6$ & \nodata \\
ad3a-22600 & $-30.463~140$ & $  0.418~588$ & \nodata  & \nodata  & \nodata  & \nodata  & $   -105.9$ & $   -105.2$ & $   -105.7$ & \nodata & $ -105.6$ \\
ad3a-22398 & $-30.963~234$ & $ -0.186~057$ & \nodata  & $    -32.7$ & \nodata  & \nodata  & \nodata  & \nodata  & \nodata  & $  -32.7$ & \nodata \\
ad3a-22823 & $-29.901~525$ & $  1.035~900$ & \nodata  & $    -80.2$ & \nodata  & \nodata  & \nodata  & \nodata  & \nodata  & $  -80.2$ & \nodata \\
ad3a-22468 & $-30.800~139$ & $ -0.021~670$ & \nodata  & $    -73.3$ & \nodata  & \nodata  & \nodata  & \nodata  & \nodata  & $  -73.3$ & \nodata \\
ad3a-22215 & $-31.420~864$ & $ -0.752~594$ & \nodata  & $    -76.0$ & \nodata  & \nodata  & \nodata  & \nodata  & \nodata  & $  -76.0$ & \nodata \\
ad3a-22267 & $-31.276~270$ & $ -0.613~094$ & \nodata  & $    -62.3$ & \nodata  & \nodata  & \nodata  & \nodata  & \nodata  & $  -62.3$ & \nodata \\
\nodata & \nodata  &    \nodata &   \nodata & \nodata  & \nodata  & \nodata  & \nodata  & \nodata & \nodata  & \nodata & \nodata\\
\enddata
\tablecomments{
    $\langle v_{\textnormal{SiO}} \rangle$ is the simple average of the velocities from each SiO detected line.
    $\langle v_{\textnormal{C}} \rangle$ is the simple average of the CS, H$^{13}$CN and H$^{13}$CO$^{+}$ detected lines. 
    All velocities are calculated in the LSRK reference frame.
    Table \ref{tab:alma_baade_velocities_cycle_2_3} can be found in its entirety in a machine-readable format.
}
\end{deluxetable*}

\section{Results} \label{sec:results}

\begin{figure*}[t]
\includegraphics[width=1.0\textwidth]{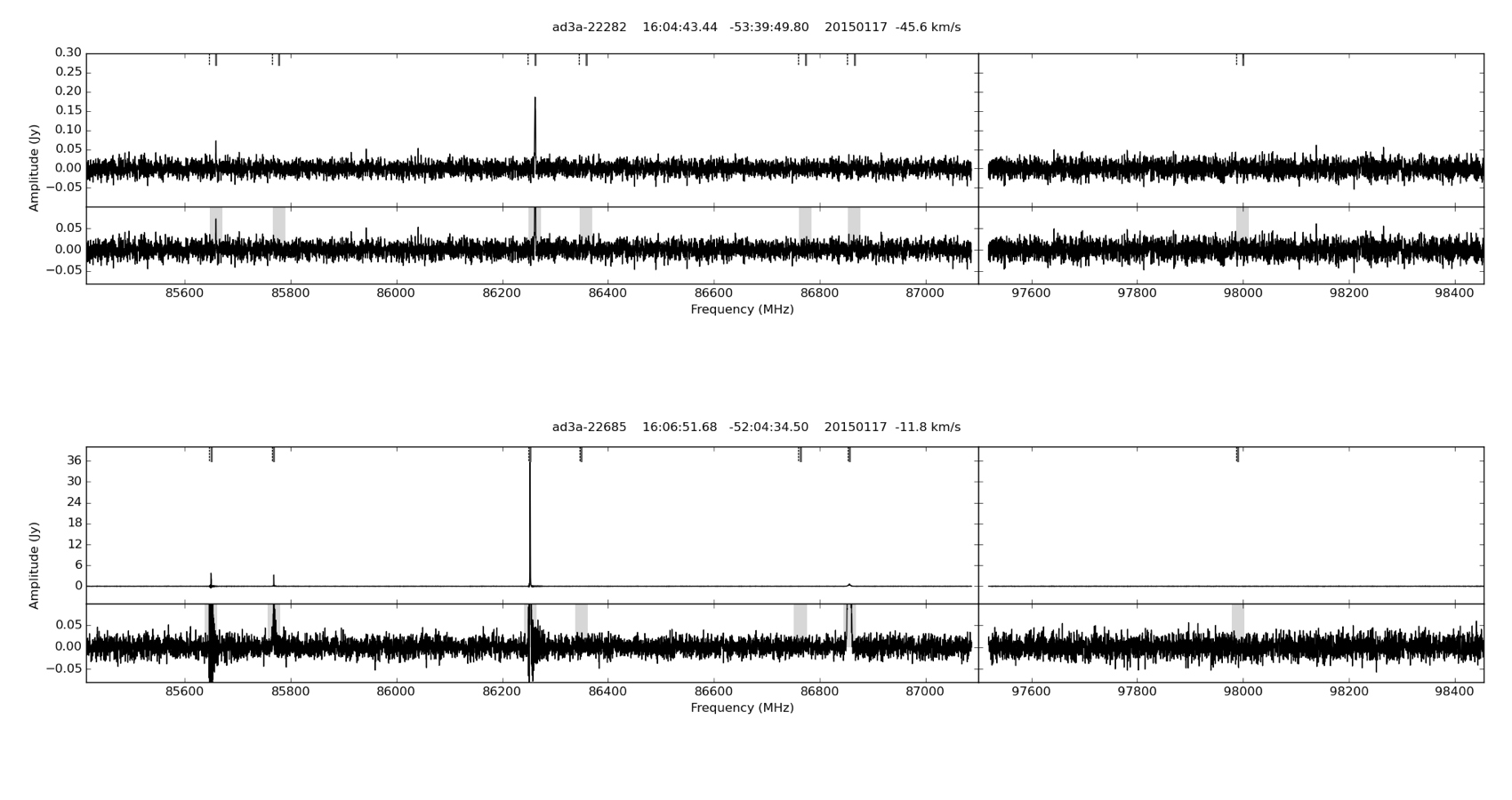}
\caption{Example spectra from SiO maser sources in the ALMA BAaDE sample. 
In the top panels, the dotted line across the top of the figures indicate where the emission would be if the source were moving at 0 km~s$^{-1}$ in the LSRK frame.
The gray lines indicate where the emission should be based on the derived velocities of the detected lines.
From left to right, the gray lines indicate SiO $v$=2, $^{29}$SiO $v$=0, SiO $v$=1, H$^{13}$CN $v$=0, H$^{13}$CO$^{+}$, SiO $v$=0 and CS.
The bottom panels are zoomed in to better identify weaker lines.
The gray regions in the bottom panels represent a $\pm50$~km~s$^{-1}$ region surrounding the region where emission should be based on the velocity derived from the detected lines.
More spectra are available at the BAaDE survey's website www.phys.unm.edu/$\sim$baade.
}
\label{fig:sio_spectra}
\end{figure*}

For each source, the line-peak brightness, $S$, or upper limit for each transition is listed in Table \ref{tab:alma_baade_detections_cycle_2_3}.
We note that SiO masers are variable, so the SiO maser brightnesses are only indicative.
Table \ref{tab:alma_baade_velocities_cycle_2_3} lists the corresponding line-peak velocities.
Two example SiO maser spectra from the ALMA BAaDE survey are presented in Figure \ref{fig:sio_spectra}.
A complete list of spectra are available at the BAaDE survey's website www.phys.unm.edu/$\sim$baade.
Section \ref{sec:detection_rates} summarizes the detection rates and trends.
Section \ref{sec:continuum_field_sources} discusses the fields containing continuum emission.

\subsection{Detection Rates} \label{sec:detection_rates}

\begin{deluxetable}{lcrrrr}[t]
\tablecaption{ALMA detection rates\label{tab_alma_det_rates}}
\tabletypesize{\scriptsize}
\tablehead{
\colhead{} & \colhead{Rest Frequency} & \colhead{Number of} & \colhead{Detection} \\
\colhead{Line Name} & \colhead{(MHz)} & \colhead{Detections} & \colhead{Rate}
}
\startdata
$\phantom{^{28}}$SiO ($J$=2--1) $v$=2 & 85~640.452 & 57 & 4.0\% \\
$^{29}$SiO ($J$=2--1) $v$=0 & 85~759.188 & 75 & 5.2\% \\
$\phantom{^{28}}$SiO ($J$=2--1) $v$=1 & 86~243.370 & 1020 & 71.2\% \\
$\phantom{^{28}}$SiO ($J$=2--1) $v$=0 & 86~846.995 & 87 & 6.1\% \\
\hline
$\phantom{^{29}}$H$^{13}$CN ($J$=1--0) $v$=0 & 86~339.920 & 12 & 0.8\% \\
$\phantom{^{29}}$H$^{13}$CO$^{+}$ ($J$=1--0) & 86~754.290 & 14 & 1.0\% \\
$\phantom{^{29}}$CS ($J$=2--1) & 97~980.950 & 50 & 3.5\% \\
\hline
$\phantom{^{29}}$Continuum emission & N/A & 4 & 0.3\%\\
\enddata
\end{deluxetable}

\begin{figure}[t]
\includegraphics[scale=0.68]{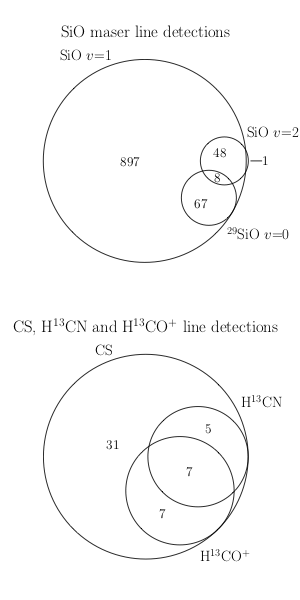}
\caption{Venn diagrams for ALMA BAaDE sources separated into SiO maser line detections (top) and carbon line detections (bottom).
All but one of the sources with SiO maser emission had SiO $v$=1 emission.
CS emission was present for all sources with H$^{13}$CN or H$^{13}$CO$^{+}$ emission.}
\label{fig:detections_venn}
\end{figure}

The detection rates for the seven main transitions in our frequency coverage are listed in Table \ref{tab_alma_det_rates}.
Our primary lines, SiO $v$=1 and CS, were detected in 1020 and 50 sources, respectively.
Three targets show emission in both lines and are discussed in Section \ref{sec:compact_cs_population}.

The Venn diagrams in Figure \ref{fig:detections_venn} illustrate how often different transitions are co-detected.
In addition to the maser line trends, 83 of the 87 sources with SiO $v$=0 emission (likely thermal) had SiO $v$=1 emission, while 5 out of the 87 stars with SiO $v$=0 emission instead show CS emission.
Thus one source, ad3a-26358, had SiO $v$=0, $v$=1 and CS emission and is a possibly associated with OH 349.39-0.01 \citep{10.1071/PH810333}.
Out of the 87 sources with SiO $v$=0 emission, 10 and 26 had SiO $v$=2 and $^{29}$SiO $v$=0 emission, respectively.
CS emission was present for all sources with H$^{13}$CN or H$^{13}$CO$^{+}$ emission.


\begin{figure*}
\includegraphics[scale=0.35]{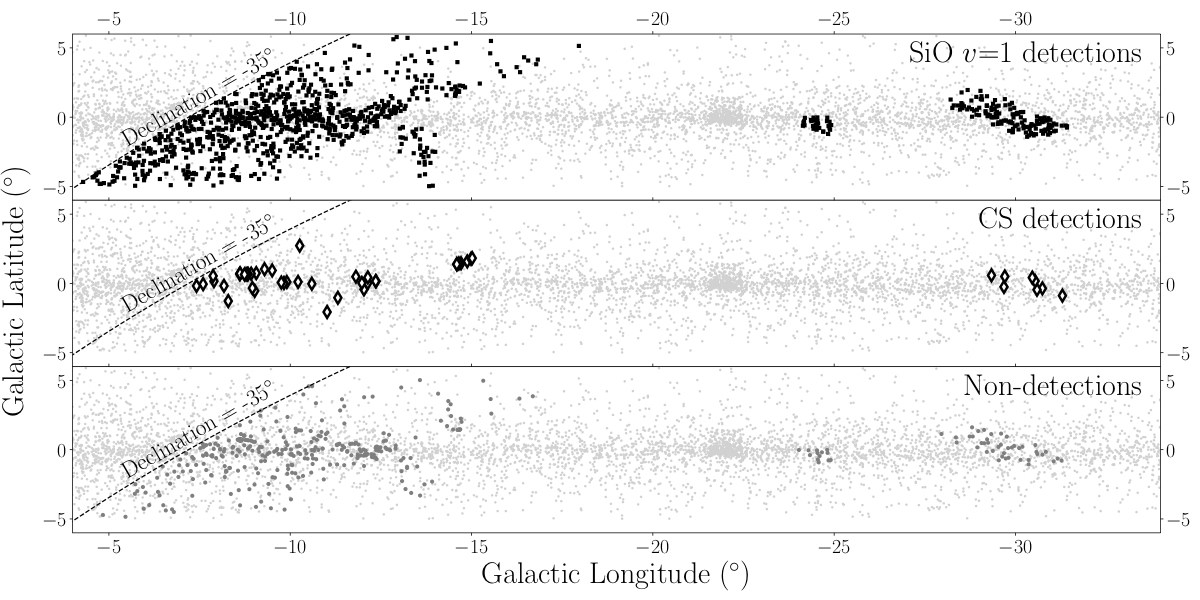}
\caption{Top, middle and bottom: SiO $v$=1 maser detections, CS detections and non-detections, respectively, from the ALMA BAaDE sample. 
The background, small gray dots represent all BAaDE targets including those still to be observed; however, the dots to the left of the $-35^\circ$ declination dashed line have been observed as part of the VLA BAaDE campaign and will be discussed in a separate paper.
No CS detections are found in the observations near $l=-25^{\circ}$.
Note that some gaps in the top-panel are due to samples that have yet to be observed.}
\label{fig_alma_detections_gal_plane}
\end{figure*}

Figure \ref{fig_alma_detections_gal_plane} shows where our observed and detected sources reside in the Galactic plane. 
Overall detection rates are consistent across Galactic longitude and latitude; however, while the CS detection rate is consistent across Galactic longitudes, they are only found very close to the Galactic plane, $|\textnormal{b}| \lesssim 1^{\circ}$.
The nature of these CS emitters is discussed in section \ref{sec:two_carbon_populations}.

\subsection{Continuum field sources} \label{sec:continuum_field_sources}

\begin{deluxetable*}{lcrrllc}
\tablecaption{Nearby continuum sources\label{tab_alma_continuum_sources}}
\tabletypesize{\scriptsize}
\tablehead{
\colhead{} & \colhead{Distance from 2MASS} & \multicolumn{2}{c}{Emission Position} &  \colhead{Possible} & \colhead{} & \colhead{Distance from}\\
\colhead{BAaDE} & \colhead{Associated Position} & \colhead{R.A.} & \colhead{Decl.} & \colhead{Continuum Emission} & \colhead{Possible Emission} & \colhead{Association}\\
\colhead{Field} & \colhead{(\arcsec)} & \colhead{(J2000)} & \colhead{(J2000)} & \colhead{Association} & \colhead{Classification} & \colhead{(\arcsec)}
}
\startdata
ad3a-22732 & 43 & 16:09:52.64 & $-$51:54:55.1 & EGO G330.95--0.18  & Outflow candidate \citep{10.1088/0004-6256/136/6/2391} & 1.1 \\
ad3a-26074 & 31 & 17:19:58.94 & $-$38:58:15.1 & G348.6972--01.0263 1 & YSO candidate \citep{10.1051/0004-6361:20077663} & 0.4 \\
ad3a-26913 & 34 & 17:20:53.47 & $-$35:47:01.9 & NGC 6334F (C) & HII region \citep{1996AAS..115...81B} & 0.7 \\
ad3a-26771 & 39 & 17:25:25.22 & $-$36:12:46.0 & GAL 351.58--00.35 & HII region \citep{1996AAS..115...81B} & 0.2 \\
\enddata
\tablecomments{
The second column is the distance between the emission and the 2MASS position for the BAaDE source.
Possible associations with the emission location were found using the SIMBAD astronomical database \citep{10.1051/aas:2000332}.
The last column indicates the distance between the emission position and the possible association.
}
\end{deluxetable*}

\begin{figure}
    \includegraphics[scale=0.15]{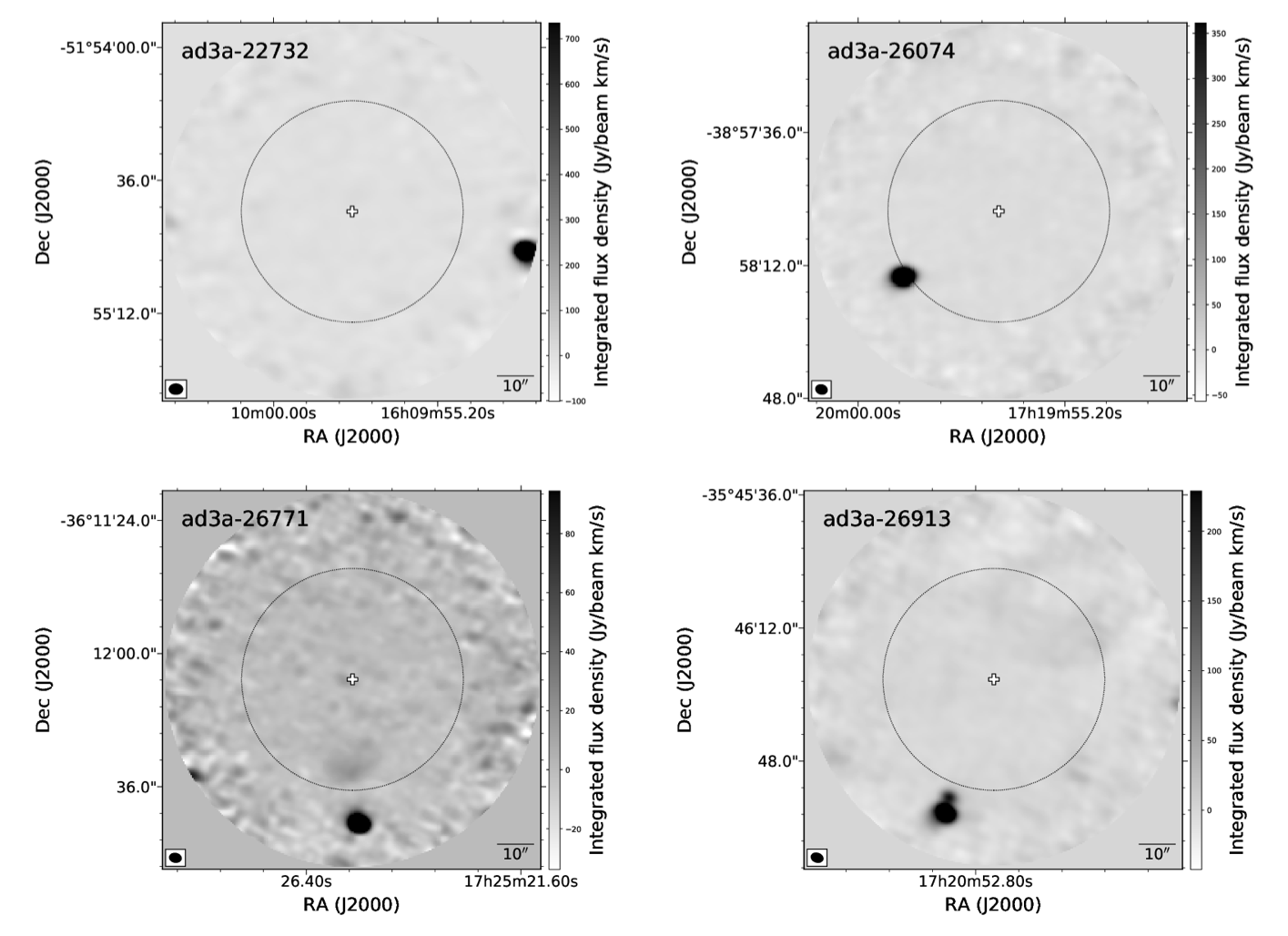}
     \caption{Integrated continuum emission (moment 0) of the continuum emission in spectral window 1 (see Table \ref{tab_alma_spw_config}) for the four fields with bright continuum sources.
     The circle represents a $30\arcsec$ distance from the 2MASS-associated BAaDE position (the center of each field).
     In each case, no line emission was found in any observed spectral window within $\approx30\arcsec$ of the 2MASS-associated BAaDE position, thus these fields are treated as non-detections.}
     \label{fig:baade_continuum_sources}
\end{figure}

Miras and supergiants can be observed as radio photospheres with continuum flux densities of a few mJy/beam for stars within a distance of $\lesssim 200$ pc \citep{10.1086/303614}, thus radio continuum emission is not expected to be detected for stars in our observations; however, continuum emission was found in four ALMA observed BAaDE fields.
In order to better understand the nature of these sources, and to determine whether the continuum emission may be associated with the BAaDE sources, the fields were imaged.

To image these fields, CASA's task \textit{tclean} was used with the H\"ogbom clean deconvolver, Briggs weighting, and Briggs robustness parameter 0.5 (equivalent to robust=0.0 in AIPS). 
The synthesized beam of the spectral window imaged at 86.25 GHz consisting of 3840 channels was approximately $3\arcsec\times2\arcsec$.

The resulting field images (Figure \ref{fig:baade_continuum_sources}) show that the continuum emission in all cases is located more than $30\arcsec$ from the BAaDE source positions, and is not associated with the targeted BAaDE sources.
The mapped continuum positions were found to be closely associated with known sources, including HII regions and Young Stellar Objects (YSO) and are detailed in Table \ref{tab_alma_continuum_sources}.
In the spectral datacubes, neither CS nor SiO emission was found at the associated 2MASS positions for the BAaDE sources in fields showing continuum sources.
Thus these sources are treated as non-detections in the following sections.


\section{Discussion} \label{sec:discussion}

The primary goal of this survey is to obtain line-of-sight velocities for evolved stars in order to map the gravitational potential and structure of the Galactic bulge and plane. 
To ensure any velocity distributions and kinematical associations are correctly interpreted, it is important to characterize the observed sample and the detections. 
First, in investigating whether the CS line would be as efficient as the SiO in determining stellar line-of-sight velocities, we found that the CS line primarily identifies young stellar objects inadvertently included in our sample through the infrared color selection, rather than tracing the evolved stars constituting the bulk of our sample (Section \ref{sec:two_carbon_populations}). 
Hence, the CS line is not an effective tracer of an AGB population and AGB stellar velocities. 
Second, for the SiO detections we examined whether individual SiO transitions provide comparable stellar line-of-sight velocities, and found that any SiO transition is suitable for determining a stellar AGB line-of-sight velocity (Section \ref{sec:velocities}).


\subsection{The CS population} \label{sec:two_carbon_populations}

\begin{figure*}[t]
\center{
\includegraphics[scale=0.60]{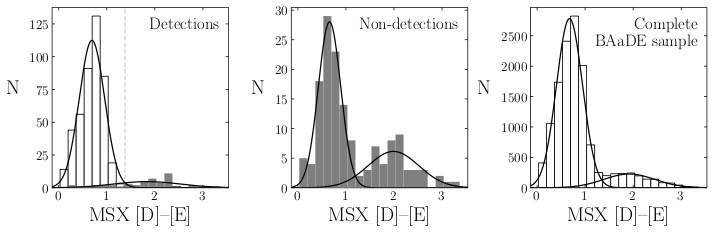}
}
\caption{Left: Zero-magnitude corrected MSX $[\textnormal{D}]-[\textnormal{E}]$ color distributions for ALMA BAaDE sources in this sample. 
The outlined bins represent sources with 86 GHz SiO $v$=1 and/or $v$=2 emission, while gray bins represent sources with CS emission. 
Separate single Gaussian distribution fits to the SiO $v$=1 \& 2 and CS distributions are over-plotted. 
The vertical gray dashed line at $[\textnormal{D}]-[\textnormal{E}]=1.38$ indicates where the overall distribution apparently changes from being dominated by SiO to CS sources.
Middle: Zero-magnitude corrected MSX $[\textnormal{D}]-[\textnormal{E}]$ color distribution for sources with no emission. 
A two-component Gaussian fit to the non-detection distribution is over-plotted.
The two-component Gaussian fit was calculated using the \textit{mclust} \citep{ISSN2073-4859_doi10.21236/ada459792} R module. 
Right: Zero-magnitude corrected MSX $[\textnormal{D}]-[\textnormal{E}]$ color distribution for the full subset of BAaDE sources with reliable MSX E-band data, where a similar bimodal distribution is found. 
The solid lines represents the best two Gaussian model fit to the full BAaDE $[\textnormal{D}]-[\textnormal{E}]$ color distribution using \textit{mclust}.}
\label{fig:o_c_detections_in_MSX_space}
\end{figure*}

The IRAS 60$\mu$m/25$\mu$m versus 25$\mu$m/12$\mu$m two-color diagram \citep{1988A&A...194..125V} discriminates between oxygen- and carbon-rich stars.
This is made possible due to the longer wavelength, 60$\mu$m band, with oxygen-rich stars tending to lie on the "evolutionary sequence" while carbon-rich stars do not. 
A degeneracy appears when mapping IRAS 2CD regions to the MSX 2CD, where the longest wavelength MSX band is centered at 21.34$\mu$m.
Thus, MSX color selections likely include a number of carbon-rich sources, some of which possibly could be detectable in the CS line similar to observations of IRC+10216 \citep{1985A&A...147..143H}.

Even though our infrared criterion primarily selects Mira-like AGB stars, it does not deselect for, e.g., supergiants and YSOs.
Separating these populations from the AGB population is important, as they are likely associated with different kinematical structures.

\subsubsection{Color distribution} \label{sec:cs_colors}

MSX color selections that include the longer MSX wavelength band $[\textnormal{E}]$ (21.3$\mu$m) separate the majority of sources with CS emission from those with SiO emission.
The left panel of Figure \ref{fig:o_c_detections_in_MSX_space} shows the distributions of SiO and CS detections as a function of MSX $[\textnormal{D}]-[\textnormal{E}]$.
The $[\textnormal{D}]-[\textnormal{E}]$ color selection provides the greatest distinction between these two populations, a separation that is also apparent in the sources without detections. 
Fitting the SiO maser and CS emission populations to individual Gaussian probability distributions with the \textit{fitdistr} function \citep[contained in the MASS \textit{R} package, see][]{ISBN0-387-95457-0}, the mean and standard deviations of the SiO emission distribution in $[\textnormal{D}]-[\textnormal{E}]$ are $0.691 \pm 0.013$ and $0.264 \pm 0.009$ magnitudes, respectively. 
Similarly, the mean and standard deviations of the Gaussian distribution fit to the CS emission distribution in $[\textnormal{D}]-[\textnormal{E}]$ are $1.84 \pm 0.10$ and $0.70 \pm 0.07$ magnitudes, respectively. 
Thus, the full distribution moves from being dominated by an SiO emission population to a CS population at $[\textnormal{D}]-[\textnormal{E}] = 1.38$. 

For sources with reliable MSX $[\textnormal{E}]$, the right panel of Figure \ref{fig:o_c_detections_in_MSX_space} shows the number distribution for the full BAaDE sample, including detections, non-detections, and sources not yet observed.
This figure indicates that the full sample shares this bimodal distribution using MSX $[\textnormal{D}]-[\textnormal{E}]$, and the majority of CS emitters are likely found in redder color sources than those typically harboring SiO masers.

\citet{10.1051/0004-6361:20040401} and \citet{10.1046/j.1365-8711.2002.05785.x} suggest that the carbon stars should be found in the bluest MSX regions (i.e. $[\textnormal{D}] - [\textnormal{E}] \lessapprox 1$), while the reddest MSX regions corresponds to PNe, YSOs, and compact HII regions.
Since nearby PNe and YSOs should show extended line emission, imaging was performed, as described in the next section, to aid in understanding the CS population.
Table \ref{tab_alma_det_rates} lists a 4\% detection rate of CS in the observed BAaDE ALMA sample, compared to 6\% of the entire BAaDE population residing in the $[\textnormal{D}]-[\textnormal{E}] > 1.38$ color region. 
However, limiting the color selection to $[\textnormal{D}]-[\textnormal{E}] > 1.38$ yields a 45\% detection rate of CS emission. 

Although the color selections using MSX E are the most useful for separating the CS and SiO detection distributions, fewer than 50\% of BAaDE targets were reliably detected in MSX E, which greatly limits its use. 
Out of the 28,062 sources in the BAaDE survey, only 13,120 sources have reliable MSX [E] detections, thus MSX color comparisons using MSX [E] cannot be applied to the entire survey population.
Additionally, only 1,600 sources, or less than 6\% of the entire sample, are in the $[\textnormal{D}]-[\textnormal{E}] > 1.38$ region.
Future work will explore using cross-matching with other missions (e.g. IRAS, \textit{Gaia}, 2MASS, AKARI, DENIS, Herschel/PACS, GLIMPSE, MIPSGAL, and WISE) to improve the color data for the full BAaDE sample.

\subsubsection{CS line images} \label{sec:cs_images}

\begin{figure*}
\includegraphics[scale=0.2]{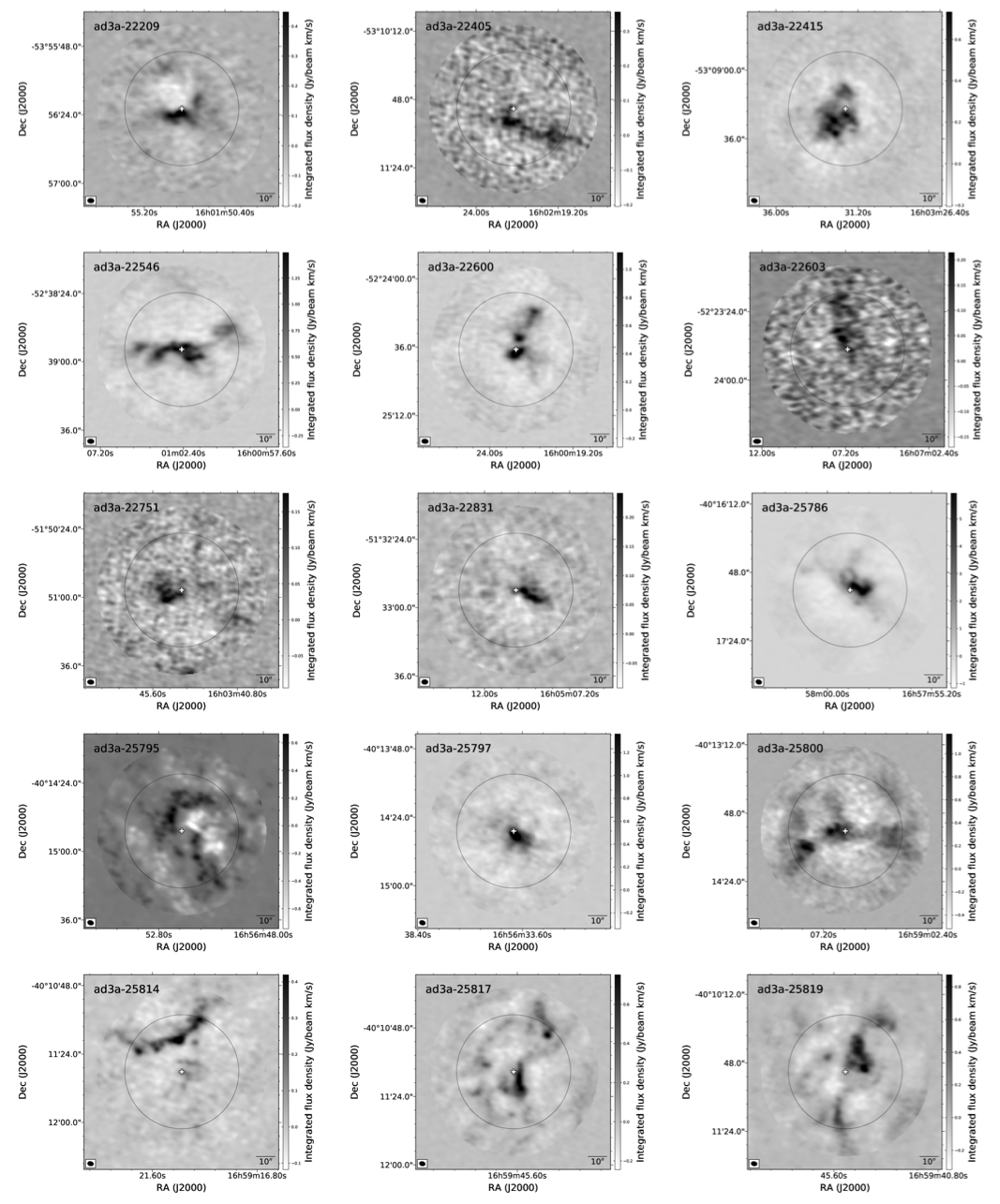}
\caption{The integrated CS flux density for ALMA BAaDE fields with CS emission (Table \ref{tab:alma_baade_detections_cycle_2_3}).
          Darker colors correspond to brighter emission.
          The flux density was integrated the FWZM of the CS emission.
          The white cross in the center of each image represents the position listed in Table \ref{tab:alma_baade_detections_cycle_2_3}.
          When SiO emission was detected in 7 of the fields, it was found to be centered on the associated 2MASS BAaDE position (see Table \ref{tab:alma_cs_emission} for details).
\label{fig:baade_alma_extended_emission}}
\end{figure*}

\begin{figure*}
\includegraphics[scale=0.2]{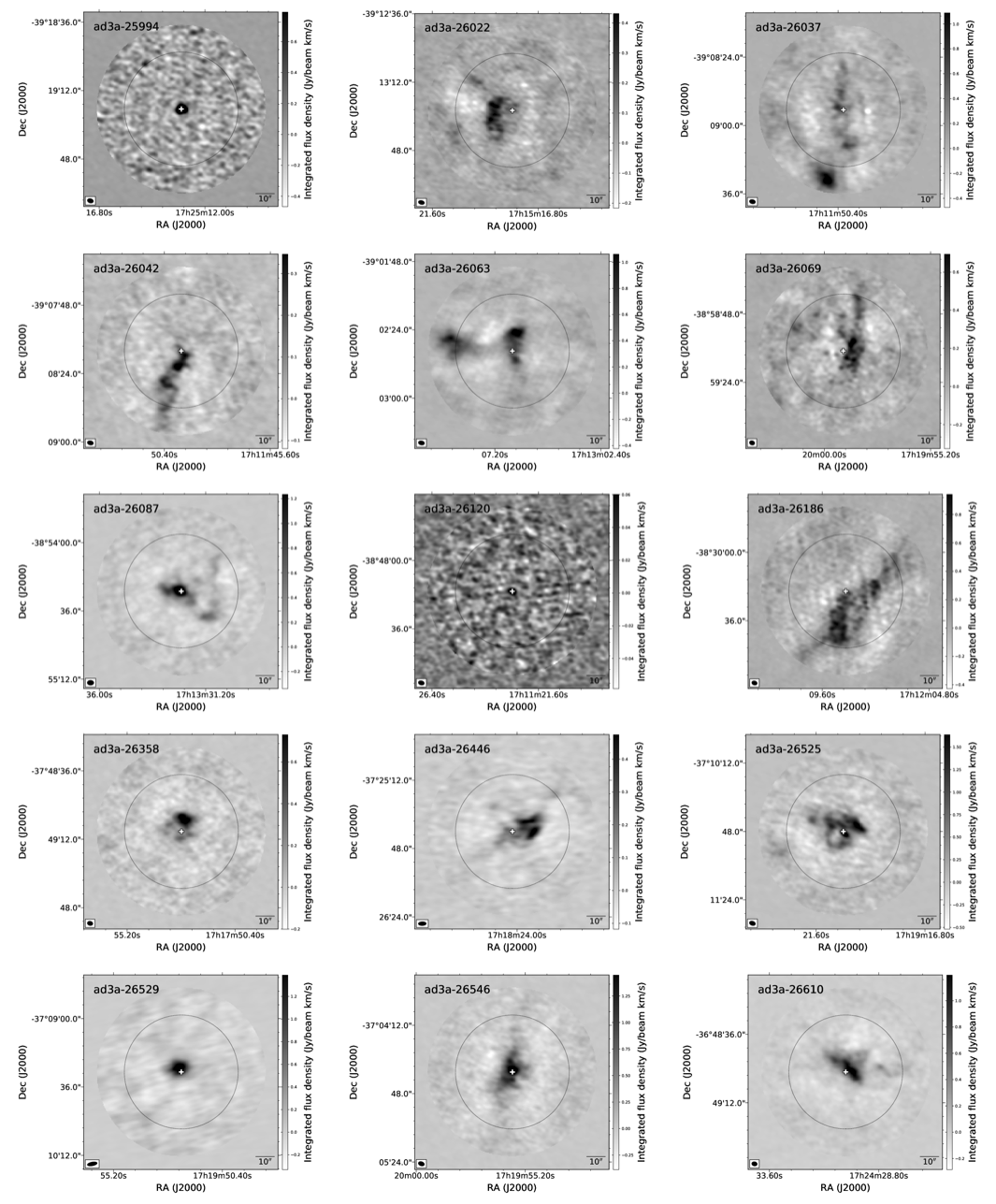}
\caption{Continued\label{fig:baade_alma_extended_emission_2}}
\end{figure*}

\begin{figure*}
\includegraphics[scale=0.2]{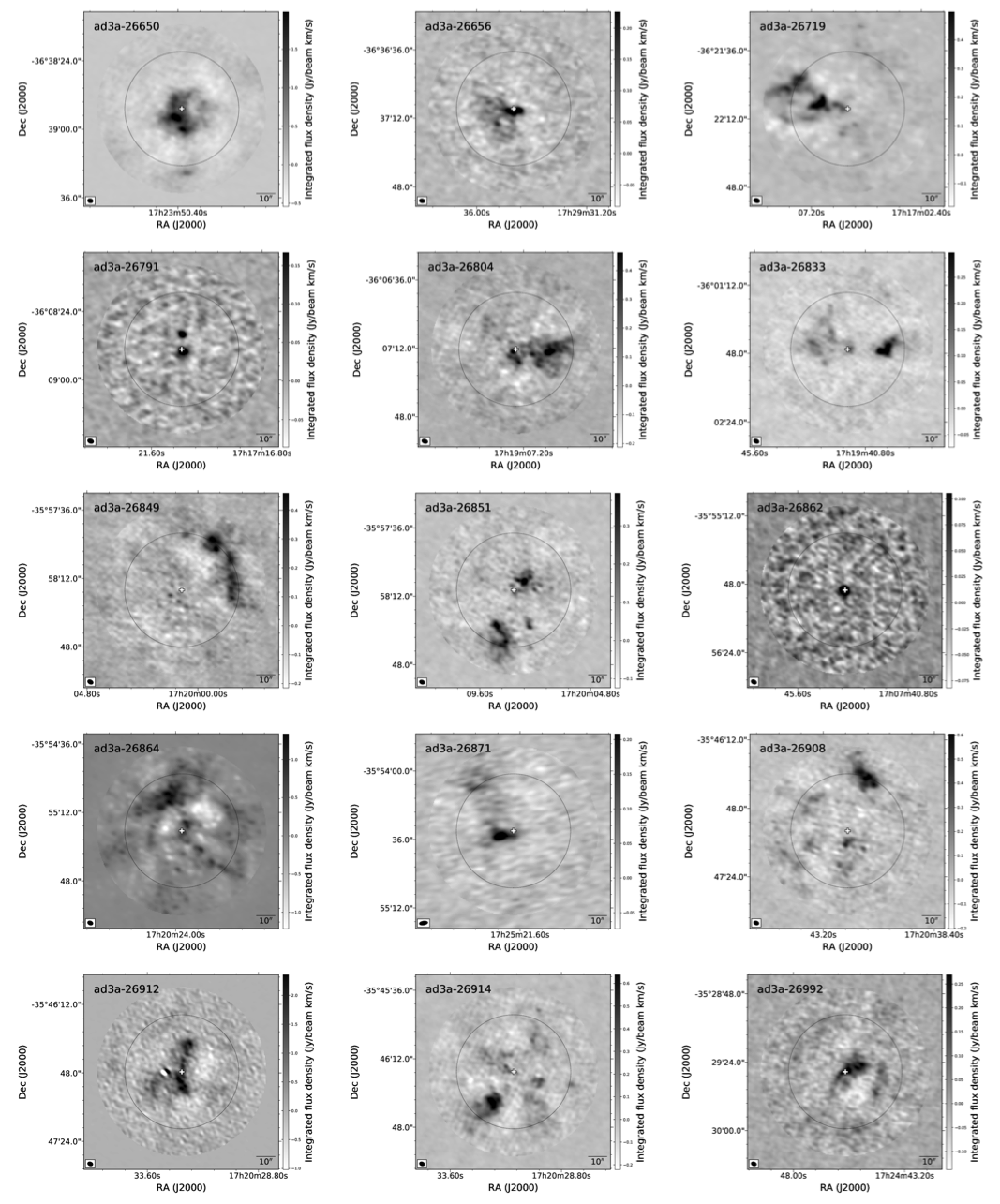}
\caption{Continued\label{fig:baade_alma_extended_emission_3}
}
\end{figure*}

\begin{figure*}
\includegraphics[scale=0.2]{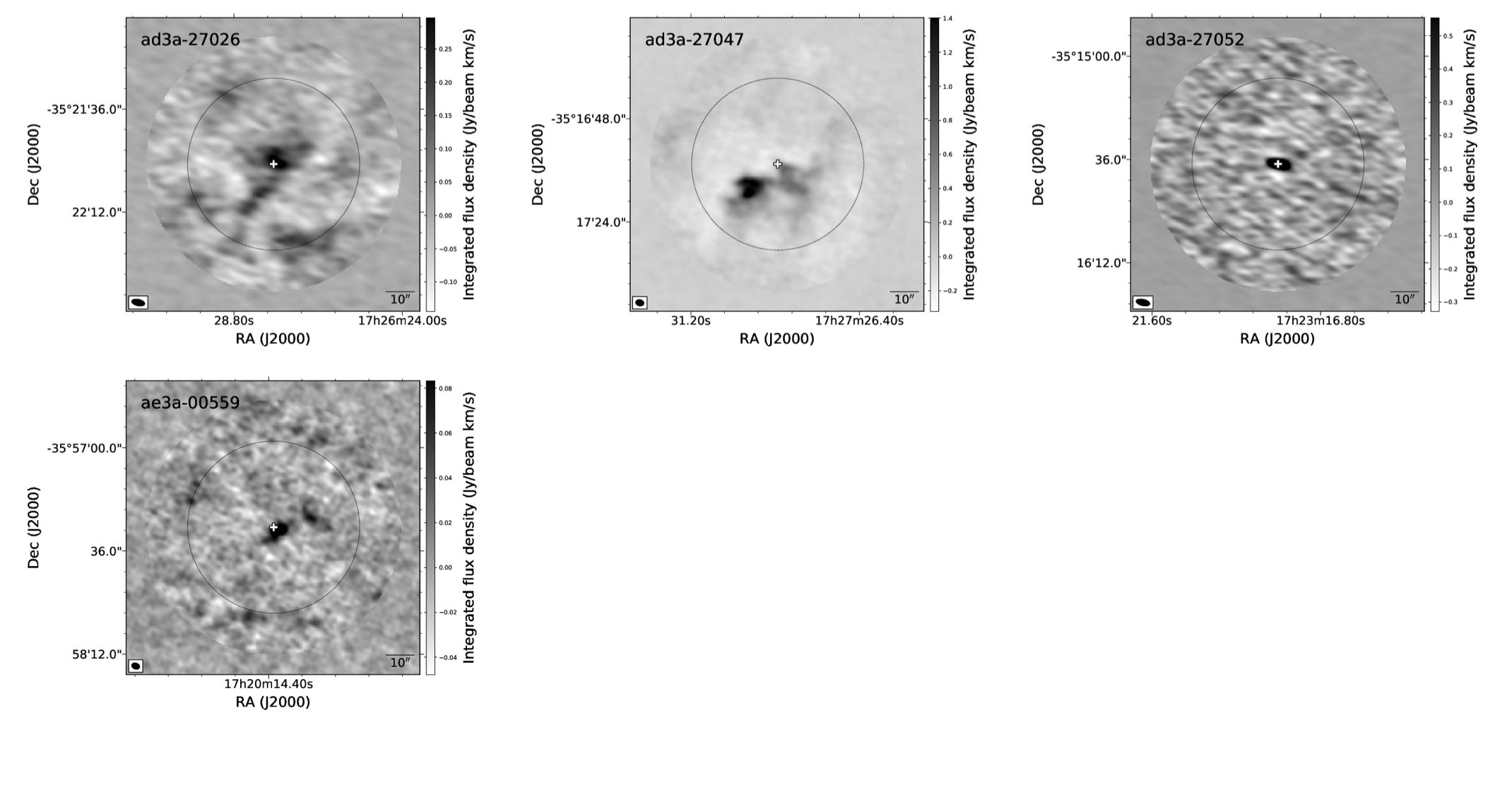}
\caption{Continued\label{fig:baade_alma_extended_emission_4}
}
\end{figure*}

\begin{longrotatetable}
\begin{deluxetable*}{lrrrrllllllll}
\tablecaption{BAaDE CS emitters and emission associations \label{tab:alma_cs_emission}}
\tabletypesize{\scriptsize}
\tablehead{
\colhead{BAaDE} & \multicolumn{2}{c}{Galactic} & 
\multicolumn{2}{c}{MSX} &
\multicolumn{6}{c}{Emission Feature} & \colhead{Possible Emission} & \colhead{Known Emission} \\
\colhead{Field} & \colhead{Longitude (\degree )} & \colhead{Latitude (\degree )} & 
\colhead{[D]--[E]} &
\colhead{[C]--[E]} &
\colhead{CS} & 
\colhead{H$^{13}$CN} &
\colhead{H$^{13}$CO+} &
\colhead{SiO $v$=0} & 
\colhead{SiO $v$=1} & 
\colhead{SiO $v$=2} &
\colhead{Association} &
\colhead{Classification}
}
\startdata
ad3a-22209 & $-31.2944$$\phantom{^{z}}$ &  $-0.8842$$\phantom{^{z}}$ &  2.34$\phantom{^{z}}$ &  2.33$\phantom{^{z}}$ & Extended & \nodata & \nodata & \nodata & \nodata  & \nodata & IRAS 15579--5347 & \\
ad3a-22405 & $-30.7420$$\phantom{^{z}}$ &  $-0.3606$$\phantom{^{z}}$ &  0.78$\phantom{^{z}}$ &  1.10$\phantom{^{z}}$ & Extended & \nodata & \nodata  & \nodata  & Compact & Compact & IRAS 15585--5302 & \\
ad3a-22415 & $-30.5935$$\phantom{^{z}}$ &  $-0.4568$$\phantom{^{z}}$ &  3.33$\phantom{^{z}}$ & \nodata$^{c}$         & Extended & Extended & Extended & \nodata & \nodata & \nodata & IRAS 15596--5301 & HII region \citep{1996AAS..115...81B} \\
ad3a-22546 & $-30.5421$$^{a}$           &  $ 0.1723$$^{a}$           &  2.36$\phantom{^{z}}$ &  2.54$\phantom{^{z}}$ & Extended & Extended & \nodata & \nodata  & \nodata & \nodata & IRAS 15572--5230 & \\
ad3a-22600 & $-30.4631$$\phantom{^{z}}$ &  $ 0.4186$$\phantom{^{z}}$ &  2.73$\phantom{^{z}}$ &  2.81$\phantom{^{z}}$ & Extended & Extended & Extended & \nodata & \nodata & \nodata & AGAL G329.537+00.422 & \\
ad3a-22603 & $-29.6826$$\phantom{^{z}}$ &  $-0.2514$$\phantom{^{z}}$ &  0.75$\phantom{^{z}}$ &  0.93$\phantom{^{z}}$ & Extended & \nodata & \nodata & \nodata  & Compact & \nodata & & \\
ad3a-22751 & $-29.7057$$\phantom{^{z}}$ &  $ 0.5024$$\phantom{^{z}}$ &  2.25$\phantom{^{z}}$ &  2.55$\phantom{^{z}}$ & Extended & \nodata & \nodata & \nodata & \nodata & \nodata & & \\
ad3a-22831 & $-29.3389$$\phantom{^{z}}$ &  $ 0.5799$$\phantom{^{z}}$ &  1.54$\phantom{^{z}}$ &  1.93$\phantom{^{z}}$ & Extended & \nodata & \nodata & \nodata & \nodata & \nodata & MWP1G330660+005800S & Infrared bubble \citep{10.1111/j.1365-2966.2012.20770.x} \\ 
ad3a-25786 & $-14.8820$$\phantom{^{z}}$ &  $ 1.5886$$\phantom{^{z}}$ &  1.98$\phantom{^{z}}$ &  1.47$\phantom{^{z}}$ & Extended & Extended & Extended & \nodata & \nodata & \nodata & & \\
ad3a-25795 & $-14.9873$$^{a}$           &  $ 1.7780$$^{a}$           &  3.05$\phantom{^{z}}$ & \nodata$^{c}$         & Extended & \nodata & \nodata & \nodata & \nodata & \nodata & J165651.7--401450& YSO candidate \citep{10.1046/j.1365-8711.1999.02890.x} \\
ad3a-25797 & $-15.0187$$\phantom{^{z}}$ &  $ 1.8252$$\phantom{^{z}}$ &  2.36$\phantom{^{z}}$ &  2.36$\phantom{^{z}}$ & Extended & Extended & Extended & \nodata & \nodata & \nodata & G344.9816+01.8252 1 \& 2 & YSO candidates \citep{10.1051/0004-6361:20077663}\\
ad3a-25800 & $-14.7102$$^{a}$           &  $ 1.4519$$^{a}$           &  1.75$\phantom{^{z}}$ &  1.13$\phantom{^{z}}$ & Extended & \nodata & Extended & \nodata & \nodata & \nodata & & \\
ad3a-25814 & $-14.6510$$^{a}$           &  $ 1.4415$$^{a}$           &  1.88$\phantom{^{z}}$ &  1.89$\phantom{^{z}}$ & Extended$^{d}$ & \nodata & \nodata & \nodata & \nodata & \nodata & & \\
ad3a-25817 & $-14.5960$$^{a}$           &  $ 1.3815$$^{a}$           &  2.41$\phantom{^{z}}$ &  2.07$\phantom{^{z}}$ & Extended & \nodata & \nodata & \nodata & \nodata & \nodata & & \\
ad3a-25819 & $-14.5933$$^{a}$           &  $ 1.3864$$^{a}$           &  2.36$\phantom{^{z}}$ &  2.48$\phantom{^{z}}$ & Extended & \nodata & \nodata & \nodata & \nodata & \nodata & & \\
ad3a-25994 & $-11.0173$$\phantom{^{z}}$ &  $-2.0634$$\phantom{^{z}}$ &  0.54$\phantom{^{z}}$ &  0.69$\phantom{^{z}}$ & Compact & Compact & \nodata & Compact & \nodata & \nodata & V1503 Sco & Star \citep{10.1086/114348} \\
ad3a-26022 & $-12.0335$$\phantom{^{z}}$ &  $-0.4319$$\phantom{^{z}}$ &  2.80$\phantom{^{z}}$ &  3.03$\phantom{^{z}}$ & Extended & \nodata & Extended & \nodata & \nodata & \nodata & GPSR 347.967--0.432 & \\
ad3a-26037 & $-12.3651$$^{a}$           &  $ 0.1566$$^{a}$           &  1.97$\phantom{^{z}}$ &  2.01$\phantom{^{z}}$ & Extended & \nodata & Extended$^{d}$ & \nodata & \nodata & \nodata & & \\
ad3a-26042 & $-12.3572$$\phantom{^{z}}$ &  $ 0.1645$$\phantom{^{z}}$ &  2.08$\phantom{^{z}}$ &  1.71$\phantom{^{z}}$ & Extended & \nodata & \nodata & \nodata & \nodata & \nodata & & \\
ad3a-26063 & $-12.1354$$^{a}$           &  $ 0.0189$$^{a}$           &  1.90$\phantom{^{z}}$ &  1.91$\phantom{^{z}}$ & Extended & \nodata & \nodata & \nodata & \nodata & \nodata& MSX5C G347.8648+00.0193 & \\
ad3a-26069 & $-11.3137$$\phantom{^{z}}$ &  $-1.0364$$\phantom{^{z}}$ &  2.03$\phantom{^{z}}$ &  2.15$\phantom{^{z}}$ & Extended & \nodata & \nodata & \nodata & \nodata & \nodata & G348.6855--01.0364 1 & YSO candidate \citep{10.1051/0004-6361:20077663} \\
ad3a-26087 & $-11.9760$$\phantom{^{z}}$ &  $ 0.0308$$\phantom{^{z}}$ &  1.18$\phantom{^{z}}$ &  1.75$\phantom{^{z}}$ & Extended & Extended & Extended & \nodata & \nodata & \nodata & AGAL G348.024+00.031 & \\
ad3a-26120 & $-12.1406$$\phantom{^{z}}$ &  $ 0.4309$$\phantom{^{z}}$ &  0.31$\phantom{^{z}}$ &  0.38$\phantom{^{z}}$ & Compact & \nodata & \nodata & Compact & \nodata & \nodata & 2MASS J17112277--3848162 & Star \citep{10.1051/0004-6361/201117874} \\
ad3a-26186 & $-11.8115$$^{a}$           &  $ 0.4871$$^{a}$           &  1.11$\phantom{^{z}}$ &  0.52$\phantom{^{z}}$ & Extended & \nodata & \nodata & \nodata & \nodata & \nodata & 2MASS J17120875--3830441 & YSO candidate \citep{10.1051/0004-6361:20077474} \\
ad3a-26358 & $-10.5959$$\phantom{^{z}}$ &  $-0.0285$$\phantom{^{z}}$ &  0.87$\phantom{^{z}}$ &  1.08$\phantom{^{z}}$ & Extended & Extended & \nodata & Compact & Compact & \nodata  & OH 349.39--0.01&OH/IR star \citep{10.1071/PH810333} \\
ad3a-26446 & $-10.2163$$\phantom{^{z}}$ &  $ 0.1123$$\phantom{^{z}}$ &  1.43$\phantom{^{z}}$ &  1.57$\phantom{^{z}}$ & Extended & \nodata & \nodata & \nodata & \nodata & \nodata & IRAS 17149--3722 &  \\
ad3a-26525 &  $-9.9067$$^{a}$           &  $ 0.1021$$^{a}$           &  2.33$\phantom{^{z}}$ &  2.44$\phantom{^{z}}$ & Extended & \nodata & \nodata & \nodata & \nodata & \nodata & & \\
ad3a-26529 &  $-9.8280$$\phantom{^{z}}$ &  $ 0.0282$$\phantom{^{z}}$ &  2.31$\phantom{^{z}}$ &  2.67$\phantom{^{z}}$ & Extended & Extended & \nodata & Compact & \nodata & \nodata & AGAL G350.172+00.029& \\
ad3a-26546 &  $-9.7548$$\phantom{^{z}}$ &  $ 0.0649$$\phantom{^{z}}$ &  1.76$\phantom{^{z}}$ &  2.11$\phantom{^{z}}$ & Extended & Extended & Extended & \nodata & \nodata & \nodata & IRAS 17165--3701 & YSO candidate \citep{10.1051/0004-6361:20020973}\\
ad3a-26610 &  $-9.0209$$\phantom{^{z}}$ &  $-0.5394$$\phantom{^{z}}$ &  2.33$\phantom{^{z}}$ &  2.32$\phantom{^{z}}$ & Extended & Extended & Extended & \nodata & \nodata & \nodata & BGPS G350.978--00.540& \\
ad3a-26650 &  $-8.9570$$^{a}$           &  $-0.3341$$^{a}$           &  2.15$\phantom{^{z}}$ &  2.46$\phantom{^{z}}$ & Extended & \nodata & Extended & \nodata & \nodata & \nodata & GPSR 351.042--0.336& \\
ad3a-26656 &  $-8.2914$$^{a}$           &  $-1.2732$$^{a}$           &  1.67$\phantom{^{z}}$ &  1.64$\phantom{^{z}}$ & Extended & \nodata & \nodata & \nodata & \nodata & \nodata & & \\
ad3a-26719 &  $-9.5025$$^{a}$           &  $ 0.9373$$^{a}$           & \nodata$^{b}$         & \nodata$^{b}$         & Extended & \nodata & \nodata & \nodata & \nodata & \nodata & & \\
ad3a-26791 &  $-9.2920$$\phantom{^{z}}$ &  $ 1.0259$$\phantom{^{z}}$ &  1.46$\phantom{^{z}}$ &  2.02$\phantom{^{z}}$ & Extended & \nodata & \nodata & \nodata & \nodata & \nodata & AGAL G350.709+01.027 & \\
ad3a-26804 &  $-9.0634$$\phantom{^{z}}$ &  $ 0.7455$$\phantom{^{z}}$ &  2.33$\phantom{^{z}}$ &  2.39$\phantom{^{z}}$ & Extended & \nodata & Extended & \nodata & \nodata & \nodata & 2MASS J17190753--3607125 & YSO candidate \citep{10.1088/0004-637X/778/2/96} \\
ad3a-26833 &  $-8.9233$$^{a}$           &  $ 0.7037$$^{a}$           &  1.53$\phantom{^{z}}$ &  1.27$\phantom{^{z}}$ & Extended & \nodata & \nodata & \nodata & \nodata & \nodata & & \\
ad3a-26849 &  $-8.8391$$^{a}$           &  $ 0.6838$$^{a}$           &  2.12$\phantom{^{z}}$ &  2.08$\phantom{^{z}}$ & Extended$^{d}$ & \nodata & \nodata & \nodata & \nodata & \nodata & 2MASS J17195986--3558075 & YSO candidate \citep{10.1088/0004-637X/778/2/96}\\
ad3a-26851 &  $-8.8226$$^{a}$           &  $ 0.6647$$^{a}$           &  2.25$\phantom{^{z}}$ &  2.04$\phantom{^{z}}$ & Extended & \nodata & \nodata & \nodata & \nodata & \nodata & & \\
ad3a-26862 & $-10.2617$$\phantom{^{z}}$ &  $ 2.7238$$\phantom{^{z}}$ &  0.97$\phantom{^{z}}$ &  1.26$\phantom{^{z}}$ & Compact & \nodata & Compact & \nodata & \nodata & \nodata & IRAS 17043--3551 & Star \citep{2018yCat.1345....0G} \\
ad3a-26864 &  $-8.7544$$^{a}$           &  $ 0.6481$$^{a}$           &  2.91$\phantom{^{z}}$ &  3.06$\phantom{^{z}}$ & Extended & \nodata & \nodata & \nodata & \nodata & \nodata & & \\
ad3a-26871 &  $-8.1734$$\phantom{^{z}}$ &  $-0.1734$$\phantom{^{z}}$ &  2.22$\phantom{^{z}}$ &  2.41$\phantom{^{z}}$ & Extended & \nodata & \nodata & \nodata & \nodata & \nodata & & \\
ad3a-26908 &  $-8.6045$$^{a}$           &  $ 0.6766$$^{a}$           &  1.73$\phantom{^{z}}$ &  1.88$\phantom{^{z}}$ & Extended & \nodata & \nodata & \nodata & \nodata & \nodata & & \\
ad3a-26912 &  $-8.6210$$\phantom{^{z}}$ &  $ 0.7064$$\phantom{^{z}}$ &  1.81$\phantom{^{z}}$ &  1.23$\phantom{^{z}}$ & Extended & \nodata & \nodata & \nodata & \nodata & \nodata & G351.380+00.707 & Dense core \citep{10.1088/0004-637X/777/2/157}\\
ad3a-26914 &  $-8.6169$$^{a}$           &  $ 0.7144$$^{a}$           &  1.85$\phantom{^{z}}$ &  1.03$\phantom{^{z}}$ & Extended & \nodata & \nodata & \nodata & \nodata & \nodata & & \\
ad3a-26992 &  $-7.8966$$^{a}$           &  $ 0.1620$$^{a}$           &  1.88$\phantom{^{z}}$ &  2.23$\phantom{^{z}}$ & Extended & \nodata & \nodata & \nodata & \nodata & \nodata & ISOGAL--P J172445.8-352927 & YSO candidate \citep{10.1051/0004-6361:20020973} \\
ad3a-27026 &  $-7.5984$$\phantom{^{z}}$ &  $-0.0539$$\phantom{^{z}}$ &  0.52$\phantom{^{z}}$ &  0.57$\phantom{^{z}}$ & Extended & \nodata & Extended & \nodata & \nodata & \nodata & 2MASS J17262767--3521552 & Star \citep{ISSN0082-4704} \\
ad3a-27047 &  $-7.4155$$^{a}$           &  $-0.1811$$^{a}$           &  2.16$\phantom{^{z}}$ &  2.10$\phantom{^{z}}$ & Extended$^{d}$ & \nodata & \nodata & \nodata & \nodata & \nodata & EGO G352.58--0.18 & YSO outflow \citep{10.1088/0067-0049/206/1/9} \\
ad3a-27052 &  $-7.8739$$\phantom{^{z}}$ &  $ 0.5385$$\phantom{^{z}}$ &  0.30$\phantom{^{z}}$ &  0.42$\phantom{^{z}}$ & Compact & Compact & \nodata & Compact & \nodata & \nodata & IRAS 17199--3512 & Star \citep{10.1086/313038} \\
ae3a-00559 &  $-8.8014$$^{a}$           &  $ 0.6543$$^{a}$           &  1.32$\phantom{^{z}}$ &  0.59$\phantom{^{z}}$ & Extended & \nodata & \nodata & \nodata & \nodata & \nodata & & \\
ae3a-00560 &  $-8.7394$$^{a}$           &  $ 0.6440$$^{a}$           &  2.03$\phantom{^{z}}$ &  1.26$\phantom{^{z}}$ & Extended$^{d}$ & \nodata & \nodata & \nodata & \nodata & \nodata & SSTGLMC G351.2586+00.6456 & YSO candidate \citep{10.1088/0004-637X/778/2/96} \\
\enddata
\tablecomments{
To simplify cross-matching between tables in this paper, the BAaDE sources are listed in order as they appeared in Tables \ref{tab:alma_baade_detections_cycle_2_3} and \ref{tab:alma_baade_velocities_cycle_2_3}.
Extended and compact comments refer to the spatial emission structure for lines in Table \ref{tab:alma_baade_detections_cycle_2_3}.
Refer to Table \ref{tab:alma_baade_detections_cycle_2_3} for upper limits for non-detected lines.
$^{29}$SiO $v$=0 is not included since the line was not detected in any of these sources.
2MASS associated positions are used unless otherwise indicated.
\tablenotetext{a}{MSX location since no 2MASS association exists.}
\tablenotetext{b}{The MSX E data is unreliable.}
\tablenotetext{c}{The MSX C data is unreliable.}
\tablenotetext{d}{The emission is offset from the BAaDE source position and may be unrelated to the MSX source.}
}
\end{deluxetable*}
\end{longrotatetable}

All fields with CS emission were imaged as described in Section \ref{sec:continuum_field_sources}. 
Figures \ref{fig:baade_alma_extended_emission}-\ref{fig:baade_alma_extended_emission_4} show the structure of the CS emission for the 50 fields where it was detected.
Table \ref{tab:alma_cs_emission} describes the spatial structure of the CS emission and the additional lines in our observing setup, as well as potential known associations.
The majority of fields show extended CS emission, thus arguing against unresolved carbon-rich stars or compact HII regions dominating the population of CS emitters. 
Coupled with the $[\textnormal{D}]-[\textnormal{E}] $ separation, this is consistent with the majority of the CS emitters not being regular AGB stars.
Consequently, the velocities derived from CS lines are, primarily, not tracing the same stellar population as the one being sampled by SiO masers.

\subsubsection{Galactic distribution} \label{sec:cs_gal_locations}

\begin{figure}
\includegraphics[scale=0.67]{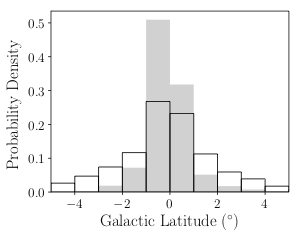}
\caption{
Probability density for the two populations from the BAaDE sample separated by $[\textnormal{D}]-[\textnormal{E}] = 1.38$. 
Sources with $[\textnormal{D}]-[\textnormal{E}] \le 1.38$ are represented in the foreground outlined bins, and sources with $[\textnormal{D}]-[\textnormal{E}] > 1.38$ are represented by the background gray bins.
The redder color sources, with larger $[\textnormal{D}]-[\textnormal{E}]$, are clustered closer to the Galactic plane than the bluer color sample.
This suggests the $[\textnormal{D}]-[\textnormal{E}] > 1.38$ population belongs to a younger stellar population that hasn't kinematically relaxed from the Galactic plane.
}
\label{fig:de_color_selection_gal_latitude}
\end{figure}

\begin{figure*}[t]
\includegraphics[width=\textwidth]{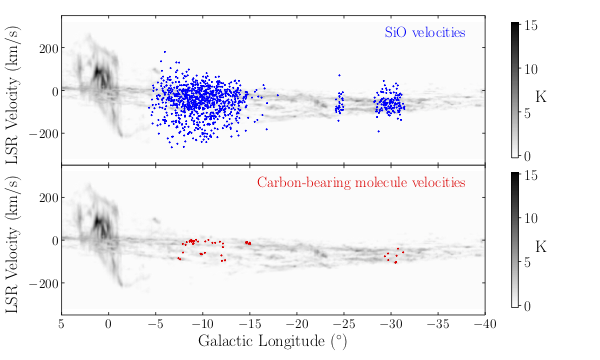}
\caption{
Galactic longitude$-$velocity (\lv) diagram for BAaDE ALMA observations from Cycles 2, 3 and 5. SiO maser derived velocities are represented in blue in the top panel, while velocities derived from CS emitters are represented by red in the bottom panel.
The velocities are taken from the corresponding $\langle \textnormal{v} \rangle$ columns in Table \ref{tab:alma_baade_velocities_cycle_2_3}. 
Integrated CO emission intensity \citep{2001ApJ...547..792D} are presented in black in both panels. 
The SiO velocities are substantially more spread out than the CO velocity range.
}
\label{fig:co_emission_comparison}
\end{figure*}

The SiO detection rates are constant across the limited range of Galactic longitudes and latitudes observed so far with ALMA.
Across Galactic latitudes, the CS detection rate increases significantly in the region within a couple of degrees of the Galactic plane. 
As noted above, the majority of CS-detected sources is characterized by $[\textnormal{D}]-[\textnormal{E}] > 1.38$.
Figure \ref{fig:de_color_selection_gal_latitude} shows the pronounced increase in such sources with Galactic latitudes, $| b | < 1^{\circ}$, a distribution very different than that of the $[\textnormal{D}]-[\textnormal{E}] \le 1.38$, SiO-dominated population.

If the $[\textnormal{D}]-[\textnormal{E}] \le 1.38$ distribution in Figure \ref{fig:de_color_selection_gal_latitude} is representative of late-type stars, then the tighter distribution of the CS dominated, $[\textnormal{D}]-[\textnormal{E}] > 1.38$ distribution, combined with the possible associations identified in Table \ref{tab:alma_cs_emission}, is likely tracing a younger population, likely coupled to the star-forming regions close to the Galactic plane.

\subsubsection{Kinematic associations} \label{sec:cs_kinematics}

The CS emission traces a much smaller overall velocity distribution than the velocities sampled by SiO, further strengthening that the CS emission traces a dynamically younger population that has not relaxed.
Figure \ref{fig:co_emission_comparison} displays the average SiO and carbon-bearing molecule velocities from Table \ref{tab:alma_baade_velocities_cycle_2_3} against Galactic longitude.
The discussion of MSX color relations in Section \ref{sec:cs_colors}, the Galactic distributions of these emitters in Section \ref{sec:cs_gal_locations}, and this \lv~ diagram suggests that the CS emitters belong to specific dynamical structures.

For the fields spanning $-15\degree < l < -5\degree$, the sources with CS emission can be split into two kinematically distinct groups with average velocities of $-10.1 \pm 1.3$ km s$^{-1}$, and $-77 \pm 5$ km s$^{-1}$. 
The first group could be associated with either the Perseus or Sagittarius arms \citep{10.1007/s10509-017-3145-5}.
The latter group has velocities placing it between the 3 kpc arm and Norma arm on an \lv~diagram \citep{10.1088/0004-637X/781/2/108, 10.3847/0004-637X/823/2/77}.
\citet{10.1007/s10509-017-3145-5} suggests sources with these velocities could also be linked with the Perseus arm.

In the $ -32\degree < l < -28\degree$ region, the CS emitters have an average velocity of $-84 \pm 8$ km s$^{-1}$ and are likely associated with the Norma arm \citep{10.1007/s10509-017-3145-5}.
These CS emitters could be used to constrain models of the spiral arms if the distances from the Sun could be determined.

%
\subsubsection{The $[\textnormal{D}]-[\textnormal{E}] \le 1.38$ CS population} \label{sec:compact_cs_population}

\begin{figure}
\includegraphics[scale=0.65]{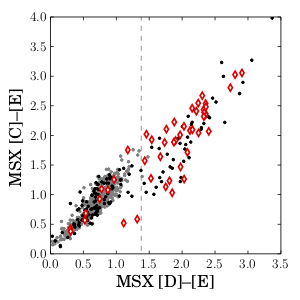}
\caption{Zero-magnitude corrected MSX $[\textnormal{C}]-[\textnormal{E}]$ versus $[\textnormal{D}]-[\textnormal{E}]$ two-color diagram for the observed ALMA BAaDE sources. 
Sources with SiO $v$=1 and/or $v$=2 emission are represented by gray dots. 
Non-detected sources are represented by black dots. 
The red diamonds represent sources with CS detections.
The gray dashed line represents the $[\textnormal{D}]-[\textnormal{E}] = 1.38$ transition between the SiO and CS dominated distributions.
The additional $[\textnormal{C}]-[\textnormal{E}]$ selection helps remove known YSOs such as ad3a-26186 from the SiO maser dominated distribution.
See Table \ref{tab:alma_cs_emission} for information on individual sources with CS emission.}
\label{fig:sio_cs_msx_everything}
\end{figure}

The vast majority of CS emitters have $[\textnormal{D}]-[\textnormal{E}] > 1.38$, but the handful of CS emitters with $[\textnormal{D}]-[\textnormal{E}] \le 1.38$ may not be YSOs, and these may instead by carbon-rich stars. 
Figure \ref{fig:sio_cs_msx_everything} demonstrates that a $[\textnormal{C}]-[\textnormal{E}]$ versus $[\textnormal{D}]-[\textnormal{E}]$ two-color diagram results in 8 sources with CS emission that cannot be separated from the region where the SiO maser sources reside.
In order of increasing $[\textnormal{D}]-[\textnormal{E}]$, ad3a-27052, ad3a-26120, ad3a-27026, ad3a-25994, and ad3a-26862 are all CS emitters with $[\textnormal{D}]-[\textnormal{E}] \le 0.97$ and have confirmed stellar associations in the literature (see Table \ref{tab:alma_cs_emission}), and are thus likely carbon-rich stars.
The remaining 3 out of the 8 sources (ad3a-22603, ad3a-22405 and ad3a-26358 in order of increasing $[\textnormal{D}]-[\textnormal{E}]$) with CS emission in the $[\textnormal{D}]-[\textnormal{E}] \le 0.87$ color region also show SiO maser emission, and may be S-type stars which typically have similar oxygen and carbon abundances.
These three sources with both SiO maser emission and CS emission are described in more detail below:


 \begin{itemize}
\item Compact, unresolved SiO $v$=1 emission is detected in ad3a-22603 centered on the 2MASS-associated BAaDE position.
The SiO $v$=1 line-peak velocity is $\approx-40$~km~s$^{-1}$.
CS emission spans the area between the 2MASS associated BAaDE source position and the location of 2MASS 16070878--5223206 (see Figure \ref{fig:baade_alma_extended_emission}).
The CS line-peak velocity is $\approx-60$~km~s$^{-1}$.

\item ad3a-22405 contains compact, unresolved SiO $v$=1 and $v$=2 emission centered on the 2MASS-associated BAaDE position.
The SiO $v$=1 and $v$=2 line-peak velocities are $\approx-60$~km~s$^{-1}$ and agree within $3$~km~s$^{-1}$. 
The CS emission is compact next to the BAaDE position, but also has a weaker emission tail extending away from the 2MASS associated BAaDE position (see Figure \ref{fig:baade_alma_extended_emission}).
The CS line-peak velocity is $\approx-40$~km~s$^{-1}$, located next to the SiO maser emission. 
The CS linear extension has a line-peak velocity of $\approx-38$~km~s$^{-1}$.
The SiO emission location is consistent with IRAS 15585--5302.

\item ad3a-26358 contains compact, and unresolved SiO $v$=0 and $v$=1, which are co-located on the 2MASS-associated BAaDE position.
The velocities of the SiO $v$=0 and $v$=1 line-peaks are $\approx 20$ km~s$^{-1}$ and differ by less than 3 km~s$^{-1}$.
Extended CS and H$^{13}$CN emission surrounds the region containing the SiO lines.
The velocities of the carbon lines are $\approx 0$ km~s$^{-1}$ and agree within 2 km~s$^{-1}$.
The 2MASS associated BAaDE location is consistent with the known OH/IR star OH 349.39--0.01 \citep{10.1071/PH810333}.
OH 349.39--0.01 has a line-of-sight velocity of 25~km~s$^{-1}$ using the mean of two OH 1612 MHz emission features separated by 26~km~s$^{-1}$ \citep{10.1071/PH810333}. 
Thus the SiO velocity is consistent with that derived from OH maser emission.

\end{itemize}

A more exhaustive treatment of the SiO and CS populations and their infrared colors can be found in M.\ Lewis et al.\ (2019), in preparation.

\subsection{The SiO maser line velocities} \label{sec:velocities}

\begin{figure*}
\includegraphics[scale=0.35]{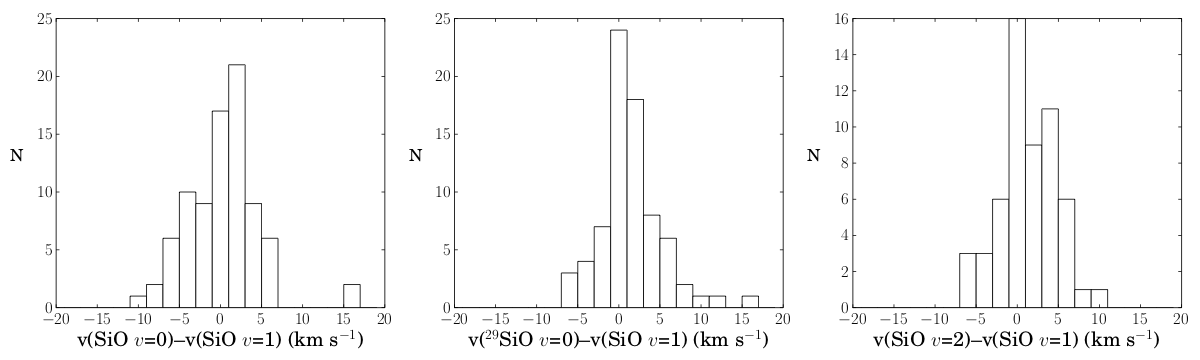}
\caption{From left to right: histogram of $\textnormal{v}(\textnormal{SiO}~v=0) - \textnormal{v}(\textnormal{SiO}~v=1)$, $\textnormal{v}(^{29}\textnormal{SiO}~v=0) - \textnormal{v}(\textnormal{SiO}~v=1)$, and $\textnormal{v}(\textnormal{SiO}~v=2) - \textnormal{v}(\textnormal{SiO}~v=1)$. On average, the SiO line-peak velocities are consistent within the channel widths.
The velocity difference distributions are also consistent the SiO $v$=1 emission velocity width distribution in Figure \ref{fig:sio_v1_FWZM}.
}
\label{fig_86_ghz_sio_delta_velocities}
\end{figure*}

Since the primary aim of the BAaDE survey is to use stars as point-mass velocity probes of the Galactic gravitational potential, it is important to ensure that no biases exist in the derived line-of-sight velocities.
Possible biases may be tested, in particular whether there are any systematic shifts in the velocity derived between the different SiO transitions. 
For instance, in Section \ref{sec:compact_cs_population} for the sources with both SiO maser and thermal CS emission, the line-peak velocities differ by $\approx$ 20~km~s$^{-1}$. 
Since the CS emission in these sources may trace outflows while the SiO traces the velocity of the central stellar source, the velocity from SiO line-peaks are more accurate probes of the AGB systemic velocity.

Figure \ref{fig_86_ghz_sio_delta_velocities} shows the distributions of velocity differences of the SiO lines relative to the SiO $v$=1 velocity. 
The Wilcoxon signed-rank test \citep{10.2307/3001968} was used to test whether the line-peak velocities differed by more than our channel resolution. 
At a 99\% significance level, there is no evidence of a systematic red- or blue-shift of any other line-peak velocity with respect to the SiO $v$=1 line, within our channel widths.
For the distributions in Figure \ref{fig_86_ghz_sio_delta_velocities}, the standard deviations are $4.6 \pm 0.3$, $3.6 \pm 0.3$ and $3.4 \pm 0.3$ km~s$^{-1}$ for the $\textnormal{v}(\textnormal{SiO}~v=0) - \textnormal{v}(\textnormal{SiO}~v=1)$, $\textnormal{v}(^{29}\textnormal{SiO}~v=0) - \textnormal{v}(\textnormal{SiO}~v=1)$, and $\textnormal{v}(\textnormal{SiO}~v=2) - \textnormal{v}(\textnormal{SiO}~v=1)$ distributions, respectively.

\begin{figure}
\includegraphics[scale=0.4]{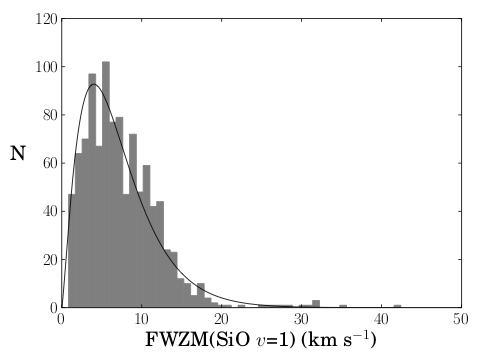}
\caption{Histogram of the full-width at zero maximum distribution for the SiO $v$=1 maser emission. 
The best-fit gamma distribution to the observed distribution with shape, $\alpha$, and rate parameters, $\beta$, of $2.3 \pm 0.10$ and $0.328 \pm 0.016$, respectively, was calculated using the \textit{fitdistr} R package, and is indicated by the solid line. 
The best-fit gamma distribution also suggests that for 90\% of the SiO $v$=1 emitters, the maser emission is limited to a 13 km~s$^{-1}$ velocity range.}
\label{fig:sio_v1_FWZM}
\end{figure}

In the context of how accurately the SiO line-peak velocity is tracing the stellar velocity, it is useful to consider the full-linewidth at zero maximum (FWZM), which yields information about the total velocity spread of the SiO emission (within the sensitivity of the observations).
A FWZM is calculated using the channels surrounding the line-peak that have emission with SNR above 5$\sigma$.  
Figure \ref{fig:sio_v1_FWZM}  displays the SiO $v$=1 emission FWZM distribution, and the best-fit gamma distribution to the observed distribution suggests that for 90\% of the sources with SiO $v$=1 emission, the emission is limited to a velocity range of 13~km~s$^{-1}$. 
This emission extent is consistent with the 12~km~s$^{-1}$ emission range of the SiO $v$=1 line suggested by the collisional pumping model of \citet{10.1051/0004-6361:20020202}, and the 15~km~s$^{-1}$ range observed by \citet{1986A&A...158...67N}.

In a study of semiregular variables (SR), \citet{10.1088/0004-6256/149/3/100} found a linear relationship between the FWZM of the 43 and 86 GHz SiO $v$=1 maser transitions, and stellar period.
The SR class contains \textbf{1)} Late-type AGB stars with V-band magnitude variations of less than 2.5 magnitudes, \textbf{2)} AGB stars with irregular periodicity, and \textbf{3)} late-type supergiants.
The International Variable Star Index \citep[VSX,][]{2006SASS...25...47W} contains periods for 61 sources with SiO $v$=1 maser emission in our observed ALMA sample.
The Spearman correlation test finds no correlation between FWZM and VSX periods with a \textit{p}-value of 0.17, thus the \citet{10.1088/0004-6256/149/3/100} relationship may not extend to Miras (typical BAaDE sources), and may point to different physical conditions for these two populations.
The long-period sample of \citet{10.1088/0004-6256/149/3/100} consists of supergiants, which follow a completely different evolutionary sequence from the AGB stars dominating our sample.
All giant semiregulars in \citet{10.1088/0004-6256/149/3/100} show a nearly flat FWZM to period relationship, and their FWZM-period trend is driven by the supergiants in their sample which had higher periods and FWZM than the rest of the sample.
Thus, the absence of a correlation between FWZM and period in this AGB-dominated sample is consistent with the giant population in \citet{10.1088/0004-6256/149/3/100}.

The FWZM for the SiO $v$=1 line represents a much larger distribution than the velocity differences in Figure \ref{fig_86_ghz_sio_delta_velocities}.
Thus, the line-peak SiO velocities derived from separate SiO transitions are consistent with the FWZM of the SiO $v$=1 emission.
The average of the SiO velocities presented in Table \ref{tab:alma_baade_velocities_cycle_2_3} represent reliable LSRK line-of-sight velocites.

The large tail in the FWZM distribution may be due to red supergiants within our source population.
\citet{10.1088/0004-6256/149/3/100} suggest that the SiO maser linewidths for supergiants stars are much larger than those found for AGB variable stars, with mean velocity widths from their fit to a Weibull distribution near 20 km~s$^{-1}$.
\citet{ISSN0004-6361} and \citet{verheyen_sio_2012} found large SiO velocity profiles (i.e. $> 20$~km~s$^{-1}$) in some supergiants in their samples.
Further, \citet{10.1088/0004-6256/149/3/100} report a general trend of the velocity widths of 43 GHz $v$=1 emission being larger than that found for 86 GHz $v$=1 emission, thus a similar or even larger tail in the FWZM distribution from the BAaDE VLA sample may be expected.


\subsubsection{Comparisons with other maser detections}

\begin{deluxetable*}{lrrll}
\tablecaption{BAaDE sources with previous SiO maser associations\label{tab:sio_velocity_comparison}}
\tabletypesize{\scriptsize}
\tablehead{
\colhead{BAaDE} & \colhead{BAaDE velocity} & \colhead{Association velocity} & \colhead{Association} & \colhead{}\\
\colhead{Source} & \colhead{(km~s$^{-1}$)} & \colhead{(km~s$^{-1}$)} & \colhead{Name} & \colhead{References}
}
\startdata
ad3a-26375 & $-173.8$ & $-174.26\phantom{0^{a}}$ & IRAS16541-3739 & \citet{1995ApJ...453..837I} \\
ad3a-26559 & $-32.4$ & $-33.69\phantom{0^{a}}$ & IRAS16560-3657 & \citet{1995ApJ...453..837I} \\
ad3a-27011 & $-49.7$ & $-52.1\phantom{00^{a}}$ & 17100-3521 & \citet{2004PASJ...56..765D} \\
ad3a-27036 & $56.9$ & $59.65\phantom{0^{a}}$ & 17197-3517 & \citet{2004PASJ...56..765D} \\
ad3a-26713 & $-47.6$ & $-46.95\phantom{0^{a}}$ & 17232-3620 & \citet{2004PASJ...56..765D} \\
ae3a-00563 & $-72.2$ & $-68.9^{a}\phantom{00}$ & 17239-3502 & \citet{2000ApJS..130..351D} \\
ad3a-26439 & $-55.6$ & $-57.8\phantom{00^{a}}$ & 17243-3724 & \citet{2000ApJS..130..351D} \\
ad3a-26558 & $65.2$ & $67.5\phantom{00^{a}}$ & 17303-3700 & \citet{2000ApJS..130..351D} \\
ad3a-27089 & $1.1$ & $-0.65\phantom{0^{a}}$ & 17331-3506 & \citet{2000ApJS..130..351D} \\
ad3a-26207 & $-81.3$ & $-80.03\phantom{0^{a}}$ & IRAS17385-3825 & \citet{1995ApJS...98..271I} \\
ad3a-26534 & $-66.5$ & $-67.83\phantom{0^{a}}$ & IRAS17414-3706 & \citet{1995ApJS...98..271I} \\
ad3a-26522 & $-148.8$ & $-149.12\phantom{0^{a}}$ & IRAS17414-3709 & \citet{1995ApJS...98..271I} \\
ad3a-26478 & $63.8$ & $64.21^{a}\phantom{0}$ & IRAS17425-3718 & \citet{1995ApJS...98..271I} \\
ad3a-26755 & $-87.7$ & $-86.095\phantom{^{a}}$ & IRAS17446-3614 & \citet{1995ApJS...98..271I} \\
ad3a-26889 & $-79.2$ & $-78.26\phantom{0^{a}}$ & IRAS17469-3550 & \citet{1995ApJS...98..271I} \\
\enddata
\tablecomments{
Unless otherwise indicated, the references listed recorded 43 GHz $v$=1 \& 2 maser detections.
The listed associated velocity is the mean from the 43 GHz $v$=1 \& 2 peak velocities.
\tablenotetext{a}{Only 43 GHz $v$=1 emission was previously detected by the reference, thus the listed velocity is from the 43 GHz $v$=1 peak.}
}
\end{deluxetable*}

\begin{deluxetable*}{lrrll}
\tablecaption{BAaDE sources with OH maser associations\label{tab:oh_velocity_comparison}}
\tabletypesize{\scriptsize}
\tablehead{
\colhead{BAaDE} & \colhead{BAaDE velocity} & \colhead{Association velocity} & \colhead{Association} & \colhead{}\\
\colhead{Source} & \colhead{(km~s$^{-1}$)} & \colhead{(km~s$^{-1}$)} & \colhead{Name} & \colhead{References}
}
\startdata
ad3a-22807 & $-56.5$ & $-57.05\phantom{0^{a}}$ & OH330.159+00.907 & \citet{1997AAS..124..509S}  \\
ad3a-26559 & $-32.4$ & $-32.2\phantom{00^{a}}$  & IRAS 16560-3657 & \citet{1991AAS...90..327T} \\
ad3a-27123 & $-31.6$ & $-31.6\phantom{00^{a}}$ & OH350.749+02.898 & \citet{1997AAS..122...79S} \\
ad3a-26025 & $-5.0$ & $-7.4\phantom{00^{a}}$ & OH347.589+00.106 & \citet{1997AAS..124..509S} \\
ad3a-26235 & $30.6$ & $30.55\phantom{0^{a}}$ & OH348.375+00.572 & \citet{1997AAS..124..509S} \\
ad3a-27011 & $-49.7$ & $-49.7\phantom{00^{a}}$ & OH350.836+02.106 & \citet{1997AAS..122...79S} \\
ce3a-00362 & $-4.7$ & $-6.65\phantom{0^{a}}$ & OH349.180+00.203	 & \citet{1997AAS..124..509S} \\
ad3a-26358 & $22.7$ & $24.75\phantom{0^{a}}$ & OH349.404-00.028 & \citet{1997AAS..124..509S} \\
ad3a-25408 & $0.0$ & $-0.8\phantom{00^{a}}$ & OH346.001-02.503 & \citet{1997AAS..124..509S} \\
ad3a-27036 & $56.9$ & $58.25\phantom{0^{a}}$ & OH352.044+00.530 & \citet{1997AAS..122...79S} \\
ad3a-26761 & $-76.1$ & $-73.85\phantom{0^{a}}$ & OH351.315-00.010 & \citet{1997AAS..122...79S} \\
ce3a-00365 & $70.4$ & $70.025^{a}$ & OH351.607+00.022, IRAS 17205-3556 & \citet{1997AAS..122...79S}, \citet{2004ApJS..155..595D}  \\
ce3a-00367 & $-12.7$ & $-13.25\phantom{0^{a}}$ & OH352.616-00.195 & \citet{1997AAS..122...79S} \\
ad3a-26439 & $-55.6$ & $-58.65^{a}\phantom{0}$ & OH350.815-01.433, IRAS 17243-3724 & \citet{1997AAS..122...79S}, \citet{1991AAS...90..327T} \\
ad3a-26731 & $-68.2$ & $-69.15\phantom{0^{a}}$ & OH352.135-01.369 & \citet{1997AAS..122...79S} \\
ad3a-26353 & $-14.6$ & $-15.9\phantom{00^{a}}$ & OH350.982-02.391 & \citet{1997AAS..122...79S} \\
\enddata
\tablecomments{
The association velocities are calculated using the simple mean between the two OH maser peaks.
Sources with only a single detected peak are not included.
\tablenotetext{a}{The association velocity is calculated using the simple average of the two values reported in the literature.}
}
\end{deluxetable*}

The velocities in Table \ref{tab:alma_baade_velocities_cycle_2_3} are consistent with the SiO and OH maser velocities in the literature.
A $4\arcsec$ search radius was used to search for possible SiO and OH maser associations in the extensive Database of Astronomical Maser Sources \citep[eDAMS,][]{10.1017/S1743921317008948}.
On average, the BAaDE SiO maser velocities differ from the SiO and OH maser association velocities by $-0.2$ and $0.4$~km~s$^{-1}$, respectively, which are smaller than the channel widths (see Table \ref{tab_alma_spw_config}).
Table \ref{tab:sio_velocity_comparison} lists the BAaDE sources with known SiO maser associations.
All OH maser associations with double peaked detections are listed in Table \ref{tab:oh_velocity_comparison}.

\section{Characteristics of the SiO masers} \label{sec:sio_maser_characteristics}

\begin{figure*}[t]
\includegraphics[scale=0.48]{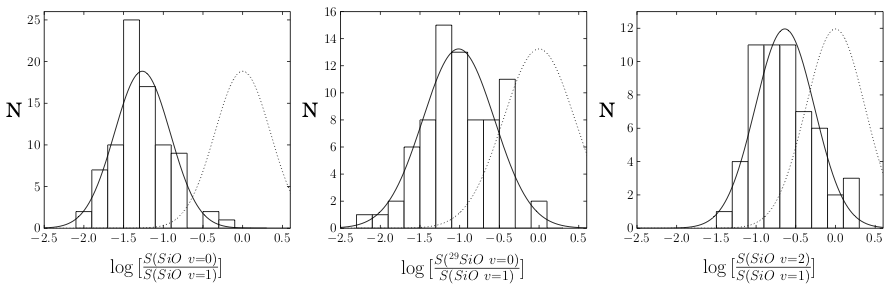}
\caption{From left to right: histograms of $\log \Big[ \frac{S(\textnormal{SiO}~v=0)}{S(\textnormal{SiO}~v=1)} \Big]$, $\log \Big[ \frac{S(^{29}\textnormal{SiO}~v=0)}{S(\textnormal{SiO}~v=1)} \Big]$ and $\log \Big[ \frac{S(\textnormal{SiO}~v=2)}{S(\textnormal{SiO}~v=1)} \Big]$ distributions, with all panels adopting a bin size of 0.2. 
The black lines represent the maximum-likelihood regression to a Gaussian distribution using \textit{fitdistr} (see Table \ref{tab:line_ratios}).
For comparison, the dotted lines indicate the location of a Gaussian distribution with the same standard deviation but with mean of 0.
SiO $v$=1 emission is, on average, the strongest line.
After SiO $v$=1, on average, the SiO $v$=2 emission is the strongest, and the SiO $v$=0 emission is the weakest.
The fit for each distribution is given in Table \ref{tab:line_ratios}.
}
\label{fig_alma_combined_sio_line_ratios}
\end{figure*}

\begin{deluxetable*}{lrrr}[t]
\tablecaption{86 GHz log distributions \label{tab:line_ratios}}
\tabletypesize{\scriptsize}
\tablehead{
\colhead{} & \colhead{} & \multicolumn{2}{c}{Log~Distribution} \\
\colhead{Transition}  & \colhead{$\langle \textnormal{SiO~transition} / v=1\rangle$} & \colhead{ Mean } & \colhead{Standard deviation}
}
\startdata
$\phantom{^{28}}$SiO $v$=0 & $ (5.4 \pm 0.5) \times10^{-2}$ & $-1.26 \pm  0.04$ & $0.35 \pm 0.02$  \\
$^{29}$SiO $v$=0                  & $ (9.1 \pm1.2 )\times10^{-2}$ &  $-1.0\phantom{4} \pm  0.5\phantom{0}$ & $0.45 \pm 0.04$ \\
$\phantom{^{28}}$SiO $v$=2  & $(2.3 \pm 0.3) \times10^{-1}$ & $-0.64 \pm  0.05$ & $0.37 \pm 0.04$ \\
\enddata
\tablecomments{ 
The second column expresses the relative brightness of the emission relative to the 86 GHz $v$=1 line based on the third column, thus the mean is the logarithm of the second column.
As an example, the SiO $v$=0 line is on average $ (5.4 \pm 0.5) \times10^{-2}$ times as bright as the $v$=1 line when both lines are detected.
The mean and standard deviations are calculated from the logarithmic distributions in Figure \ref{fig_alma_combined_sio_line_ratios}.
Using the Anderson--Darling normality test, at a 99\% significance level, all logarithmic distributions are consistent with Gaussian distributions.}
\end{deluxetable*}

This sample represents the largest SiO maser population observed at 86 GHz using the same receivers and spectral setup.
Thus this partial ALMA BAaDE sample offers a unique opportunity to unravel the nature of the pumping mechanisms and environments that give rise to this emission. 
In Section \ref{sec:sio_v1} we discuss the characteristics of the SiO $v$=1 emission.
Section \ref{sec:sio_v0} briefly explores the simultaneous relative line strength of the thermal SiO $v$=0 line to that of the primary maser line in our sample, SiO $v$=1.
Sections \ref{sec:sio_29} and \ref{sec:sio_v2} consider both the detection rates and relative line strengths of these lines and discuss these results in the context of current pumping models.

In order to remain consistent with previous comparisons in the literature, we have adopted the standard of placing the 86 GHz SiO $v$=1 line in the denominator and other SiO lines in the numerator when considering line ratios. 
We will also use logarithmic line ratios which are unbiased to the choice of numerator when comparing two quantities.

\subsection{The 86 GHz SiO $v$=1 maser line} \label{sec:sio_v1}

Table \ref{tab_alma_det_rates} notes a 71.2\% detection rate for SiO in the ALMA BAaDE sample, but the detection rate jumps to 80\% if limiting source selection to $[\textnormal{D}]-[\textnormal{E}] \le 1.38$, with only  a 2\% chance of SiO maser emission expected in the $[\textnormal{D}]-[\textnormal{E}] > 1.38$ color region.
The non-detection rate in the $[\textnormal{D}]-[\textnormal{E}] \le 1.38$ color region is likely dominated by source variability and sensitivity; however, carbon-rich stars should also reside in this oxygen-rich region and were discussed in Section \ref{sec:compact_cs_population}. 
\citet{10.3847/1538-4357/aaccf3} re-detected 66 out of 86 (77\%) known BAaDE SiO maser sources at 86 GHz, and \citet{10.3847/1538-4357/aae77d} re-detected 26 out of 33 (79\%) known BAaDE SiO maser sources at 43 GHz. 
Both detection rates are consistent with an 80\% random detection rate, and suggest that source variability is the dominant contributor to the SiO non-detection rate, and may indicate that at some point all oxygen-rich BAaDE sources will easily show 86 GHz SiO $v$=1 emission. 

As discussed in Section \ref{sec:cs_colors}, almost all SiO $v$=1 emitters are found in the $[\textnormal{D}]-[\textnormal{E}] \le 1.38$ color region.
Similar MSX color selections can be made in $[\textnormal{A}]-[\textnormal{E}]$ and $[\textnormal{C}]-[\textnormal{E}]$, although $[\textnormal{D}]-[\textnormal{E}]$ color is superior at separating the SiO $v$=1 and CS emitting populations.
No other correlations are found between SiO $v$=1 detection rates or emission strengths relative to the six MSX colors (e.g. $[\textnormal{A}]-[\textnormal{C}]$, $[\textnormal{A}]-[\textnormal{D}]$, $[\textnormal{A}]-[\textnormal{E}]$, $[\textnormal{C}]-[\textnormal{D}]$, $[\textnormal{C}]-[\textnormal{E}]$, and $[\textnormal{D}]-[\textnormal{E}]$).

\subsection{The thermal 86 GHz SiO $v$=0 lines} \label{sec:sio_v0}
SiO $v$=0 emission was detected in 87 sources with 83 of the sources having SiO $v$=1 emission.
The left panel of Figure \ref{fig_alma_combined_sio_line_ratios}, displays the distribution of line-peak ratios, plotted as number versus $\log{\bigg[ \frac{\textnormal{S(SiO~}v=0)}{\textnormal{S(SiO~}~v=1)} \bigg]}$, for sources where both SiO $v$=0 and 1 were detected.
On average, for sources in which both lines are detected, the SiO $v$=0 emission is 5\% as bright as the SiO $v$=1 maser emission (see Table \ref{tab:line_ratios}).
The symmetry in the source distribution and the absence of a secondary, higher line-ratio component suggests that no significant SiO $v$=0 \textit{maser} sample is present, thus differentiating the 86 GHz sample from the 43 GHz $v$=0 lines which sometimes indicate maser emission \citep{10.1086/386541, 10.1051/0004-6361/201527174}.
No correlations between SiO $v$=0 detection rates or emission strengths were found with respect to any MSX colors.

\subsection{The $^{29}$SiO $v$=0 maser and a higher density regime} \label{sec:sio_29}
$^{29}$SiO $v$=0 emission was detected in 75 sources, all of which had detectable SiO $v$=1 emission. 
The distribution of line ratios in the center panel of Figure \ref{fig_alma_combined_sio_line_ratios} implies that on average, the $^{29}$SiO $v$=0 emission brightness will only be 9\% of the brightness of the SiO $v$=1 emission (see Table \ref{tab:line_ratios}). 
A comparison between left and center panels of Figure \ref{fig_alma_combined_sio_line_ratios} suggests that at the same sensitivity, the detection rate of $^{29}$SiO $v$=0 should be higher than that of SiO $v$=0; however, the lower detection rate is possibly due to the Milky Way $^{29}$SiO abundance being about 6\% of that of $^{28}$SiO \citep{10.3847/1538-4357/aa67e6}, or the $^{29}$SiO $v$=0 pumping is possibly less efficient than that of SiO $v$=1.

As discussed in Section \ref{sec:cs_colors}, the SiO $v$=1 emitters are primarily found in the MSX color region defined by $[\textnormal{D}]-[\textnormal{E}] \le 1.38$, thus sources with both SiO $v$=1 and $^{29}$SiO $v$=0 are found in this same color region.
Within this color region, the sources with $\log \Big[ \frac{S(^{29}\textnormal{SiO}~v=0)}{S(\textnormal{SiO}~v=1)} \Big] >-0.55$ all reside in the $[\textnormal{D}]-[\textnormal{E}] \ge 0.68$ color region which contains $\approx 50\%$ of all SiO $v$=1 emitters.
Similarly, using any of the other 5 MSX colors, the $\log \Big[ \frac{S(^{29}\textnormal{SiO}~v=0)}{S(\textnormal{SiO}~v=1)} \Big] >-0.55$ emitters reside in color regions containing the reddest $\approx 50\%$ of all SiO $v$=1 emitters.
The redder MSX color sub-regions, within the $[\textnormal{D}]-[\textnormal{E}] \le 1.38$ color region, may indicate that higher densities in the circumstellar envelope are more conducive to $^{29}$SiO $v$=0 emission.
If so, we may expect to find a predominance of the brighter $^{29}$SiO $v$=0 emitters to be redder, which was not observed.
The $^{29}$SiO $v$=0 emission is not particularly bright for sources with $\log \Big[ \frac{S(^{29}\textnormal{SiO}~v=0)}{S(\textnormal{SiO}~v=1)} \Big] >-0.55$.
Instead, this may be explained by weaker SiO $v$=1 emission.

Assuming these are oxygen-rich stars, \citet{2014A&A...565A.127D} suggest that at higher densities, the 86 GHz SiO $v$=1 maser becomes quenched and the SiO $v$=3 emission dominates at 86 GHz. 
Additionally, the 43 GHz $v$=1, 2 and 3 emission should all be very strong and detectable at these higher densities. 
Thus observations of the redder MSX regions at both 43 and 86 GHz could help determine whether the 86 GHz SiO $v$=1 transition is being quenched, and whether the $^{29}$SiO $v$=0 emission is relatively brighter in those cases as a result.
No other correlation is found between $\log \Big[ \frac{S(^{29}\textnormal{SiO}~v=0)}{S(\textnormal{SiO}~v=1)} \Big]$ and any of the six possible MSX colors.

\subsection{The 86 GHz SiO $v$=2 lines} \label{sec:sio_v2}

Of the observed SiO lines, the SiO $v$=2 has the lowest detection rate.
Only one source with SiO $v$=2 emission had no $v$=1 emission. 
This instantaneous 4\% detection rate (Table \ref{tab_alma_det_rates}) is consistent with the 4.7\% detection rate of this transition by \citet{10.3847/1538-4357/aaccf3}, who sampled known SiO emitting sources from the BAaDE survey. 

The right panel of Figure \ref{fig_alma_combined_sio_line_ratios} shows the distribution of line-peak ratios, plotted as number versus $\log \Big[ \frac{S(\textnormal{SiO}~v=2)}{S(\textnormal{SiO}~v=1)} \Big]$.
Table \ref{tab:line_ratios} indicates that the SiO $v$=2 emission is 23\% as bright as the SiO $v$=1 line when both SiO $v$=1 and $v$=2 emission are detected.
Even though $\log \Big[ \frac{S(\textnormal{SiO}~v=2)}{S(\textnormal{SiO}~v=1)} \Big]$ is on average larger than the other logarithmic line ratios discussed above, SiO $v$=2 has the lowest detection rate.
\citet{1996A&A...314..883B} proposed that strong 86 GHz SiO $v$=2 maser emission is associated with S-type stars, if so, then the majority of our sources where the 86 GHz SiO $v$=2 transition was not detected are likely associated with Mira-type oxygen-rich stars.
The anomalous weakness in the 86 GHz SiO $v$=2 line strength has been observed since the 1970s and an explanation for it was first proposed by \citet{1981ApJ...247L..81O}, who noted the proximity between the 8$\mu$m transitions of $v$=1--2 $J$=0--1 SiO \citep{1974ApJ...191L..37G} and $v$=0--1 $J$=12$_{7,5}$--11$_{6,6}$ para-H$_2$O transition \citep{1952_benedict}.
Thus H$_2$O emission transfers energy into the SiO $v$=2 $J$=1 state, and the relative overpopulation in the SiO $v$=2 $J$=2 state is removed.
All of the sources with SiO $v$=2 emission may be S-type stars, where the effect of the H$_2$O line overlap is minimized and thus where strong SiO $v$=2 emission is possible.

The radiative modeling of \citet{2014A&A...565A.127D} suggests that $\frac{S(\textnormal{SiO}~v=2)}{S(\textnormal{SiO}~v=1)}$ may trace the density in the circumstellar envelope with SiO $v$=1, 2, \& 3 maser emission peaking at successively higher densities.
The denser circumstellar shells are found in the redder MSX color-color regions and may include the densities required for the brighter SiO $v$=2 emission.
The small sample size and narrow MSX color range may explain the lack of correlation found between these line ratios and MSX colors.
Alternatively, since these sources are variable, density changes throughout the stellar cycle could alter the line ratios.
In this case, the $\frac{S(\textnormal{SiO}~v=2)}{S(\textnormal{SiO}~v=1)}$ should be maximized during time periods of higher densities in the pulsating circumstellar envelope.
Completing the full ALMA BAaDE survey can lead to a more thorough understanding of the circumstellar densities in this sample.

\section{Summary}
The BAaDE survey utilizes ALMA observations of 86 GHz SiO masers to trace stellar line-of-sight velocities in regions that the VLA cannot observe, more specifically the far side of the Galactic bulge.
Since no statistically significant bias is found between the line-peak velocities of the 86 GHz SiO $v$=0, 1 \& 2 and $^{29}$SiO $v$=0 lines, if only one SiO line is detected (almost always the SiO $v$=1 line), its line-peak velocity should be sufficient as a qualitative stellar line-of-sight velocity probe.
SiO $v$=1 emission is detected in 71\% of sources in the survey and is detected in the vast majority of sources with other SiO line detections.

The observed SiO $v$=1 emission is likely made-up of multiple separate masing regions in the circumstellar envelope.
We find that the SiO $v$=1 emission indicates that in 90\% of the cases, the SiO emission is found within a velocity range of 13 km~s$^{-1}$ around the anticipated stellar velocity.

The CS $J=2-1$ line at 98 GHz was also observed, and was detected in less than 4\% of the observed sources. 
A demarcation can be made between the majority of SiO sources residing in color regions with MSX $[\textnormal{D}]-[\textnormal{E}] \le 1.38$ and CS emitters dominating the color population with $[\textnormal{D}]-[\textnormal{E}] > 1.38$.
CS emission is found in approximately 45\% of the sources with $[\textnormal{D}]-[\textnormal{E}] > 1.38$, while the detection rate for SiO $v$=1 in the $[\textnormal{D}]-[\textnormal{E}] \le 1.38$ region is approximately 80\%.
There is evidence that the CS emission is predominantly detected in a different population than late-type stars.
In particular, we find the following:

\begin{itemize}
\item[1.] The majority of CS emitters are found in the MSX color region defined by $[\textnormal{D}]-[\textnormal{E}] > 1.38$. 
Conversely, all SiO masers are found in the $[\textnormal{D}]-[\textnormal{E}] \le 1.38$ color region.
Additionally, \citet{10.1051/0004-6361:20040401} and \citet{10.1046/j.1365-8711.2002.05785.x} have suggested that carbon-stars should reside in the $[\textnormal{D}]-[\textnormal{E}] \lesssim 1$ region, and the redder. $[\textnormal{D}]-[\textnormal{E}] \gtrsim 1$, color regions are more likely to contain YSOs, PNe and compact HII regions.
Thus the CS emitters in the $[\textnormal{D}]-[\textnormal{E}] > 1.38$ color regions may not be associated with late-type stars.

\item[2.] The CS emission is largely extended, thus the majority are likely not associated with compact HII regions.
10 of the CS emitters may be associated with YSOs (Table \ref{tab:alma_cs_emission}).

\item[3.] The MSX point source catalog covered the full Galactic plane with $|b| < 5^{\circ}$, but the $[\textnormal{D}]-[\textnormal{E}] > 1.38$ population is mostly found in $|b| < 1^{\circ}$.
This strengthens the suggestion that the $[\textnormal{D}]-[\textnormal{E}] > 1.38$ population is young and may be coupled to the star-forming regions close to the Galactic plane.

\item[4.] The \lv~ diagrams (Figure \ref{fig:co_emission_comparison}) show that the sources with carbon-bearing molecular transitions may be tied to specific dynamical structures.
The locations of sources with carbon-bearing molecular transitions in the \lv~ diagrams are consistent with the locations of spiral arms \citep{10.1088/0004-637X/781/2/108, 10.3847/0004-637X/823/2/77,10.1007/s10509-017-3145-5}.
\end{itemize}

Thus the CS emission may be tied to YSOs, but we cannot completely rule out PNe.
Since CS emission may be associated with outflows of a different population, the SiO maser emission is a superior velocity tracer of the evolved stellar population.

\acknowledgments
This paper makes use of the following ALMA data: ADS/JAO.ALMA\#2013.1.01180.S, \\ADS/JAO.ALMA\#2015.1.01289.S, and \\ADS/JAO.ALMA\#2017.1.01077.S.
ALMA is a partnership of ESO (representing its member states), NSF (USA) and NINS (Japan), together with NRC (Canada), MOST and ASIAA (Taiwan), and KASI (Republic of Korea), in cooperation with the Republic of Chile. The Joint ALMA Observatory is operated by ESO, AUI/NRAO and NAOJ.

Support for this work was provided by the NSF through award SOSP A2-025 from the NRAO.

Support for this work was provided by the NSF through the Grote Reber Fellowship Program administered by Associated Universities, Inc./National Radio Astronomy Observatory.

This material is based upon work supported by the National Science Foundation under Grant Number 1517970 to UNM and 1518271 to UCLA. Any opinions, findings, and conclusions or recommendations expressed in this material are those of the authors and do not necessarily reflect the views of the National Science Foundation.

The National Radio Astronomy Observatory is a facility of the National Science Foundation operated under cooperative agreement by Associated Universities, Inc.

This research has made use of the SIMBAD database, operated at CDS, Strasbourg, France.

\facility{ALMA}
\software{CASA \citep{2007ASPC..376..127M}, \textit{fitdistr} \citep{ISBN0-387-95457-0}, MASS \citep{ISBN0-387-95457-0}, \textit{mclust} \citep{ISSN2073-4859_doi10.21236/ada459792}}

\end{document}